\documentclass[dvipdfmx]{ptephy}

\usepackage{amsfonts}
\usepackage{amssymb}
\usepackage{amsmath}

\usepackage{ulem}
\usepackage{color}
\newcommand{\Slash}[1]{{\ooalign{\hfil/\hfil\crcr$#1$}}}

\begin{document}

\title{Lattice QCD analysis of the Polyakov loop 
  in terms of Dirac eigenmodes
}

\author{
  \name{\fname{Takumi} \surname{Iritani}}{1},
  \name{\fname{Hideo}  \surname{Suganuma}}{2},
}

\address{
  \affil{1}
  {High Energy Accelerator Research Organization (KEK),
  1-1 Oho, Tsukuba, Ibaraki 305-0801, Japan \\ \email{iritani@post.kek.jp}}
  \affil{2}
  {Department of Physics, Graduate School of Science,
    Kyoto University, 
  Kitashirakawa-oiwake, Sakyo, Kyoto 606-8502, Japan \\
  \email{suganuma@ruby.scphys.kyoto-u.ac.jp}
}}

\begin{abstract}
  Using the Dirac-mode expansion method, which keeps the gauge invariance, 
  we analyze the Polyakov loop in terms of the Dirac modes 
  in SU(3) quenched lattice QCD in both confined and deconfined phases.
  First, to investigate the direct correspondence 
  between confinement and chiral symmetry breaking, 
  we remove low-lying Dirac-modes from the confined vacuum 
  generated by lattice QCD.
  In this system without low-lying Dirac modes, 
  while the chiral condensate $\langle \bar{q} q\rangle$ is extremely reduced, 
  we find that the Polyakov loop is almost zero and $Z_3$-center symmetry 
  is unbroken, which indicates quark confinement.
  We also investigate the removal of ultraviolet (UV) Dirac-modes, 
  and find that the Polyakov loop is almost zero.
  Second, we deal with the deconfined phase above $T_c$, and 
  find that the behaviors of the Polyakov loop and $Z_3$-symmetry are 
  not changed without low-lying or UV Dirac-modes.
  Finally, we develop a new method to remove low-lying Dirac modes 
  from the Polyakov loop for a larger lattice of $12^3 \times 4$ 
  at finite temperature, and find almost the same results. 
  These results suggest that 
  each eigenmode  has the information of confinement, i.e.,
  the ``seed''  of confinement
  is distributed in a wider region of the Dirac eigenmodes
  unlike chiral symmetry breaking,
  and there is no direct correspondence between
  confinement and chiral symmetry breaking through Dirac-eigenmodes.
\end{abstract}

\subjectindex{B02, B03, B64}



\maketitle
\section{Introduction}

Quantum chromodynamics (QCD) has been established 
as the fundamental theory of the strong interaction. 
However, its non-perturbative phenomena 
such as color confinement and chiral symmetry breaking \cite{Nambu:1961} 
are not yet fully understood. 
It is an intriguing subject to clarify the relation between 
these non-perturbative phenomena \cite{Suganuma:1995DGL,
  Miyamura:1995,Woloshyn:1995,Gattringer:2006,Bruckmann:2007,
  Bilgici:2008a,Bilgici:2008b,Synatschke:2008,
  Gattner:2005,Hollwieser:2008,
  Kovacs:2010,Lang:2011,Glozman:2012,
  Suganuma:2011,Gongyo:2012,Iritani:2012,TMDoi}.

As for a possible evidence on the close relation between 
confinement and chiral symmetry breaking, 
lattice QCD calculations have shown 
the almost simultaneous chiral and deconfinement transitions 
at finite temperature and in finite volume \cite{Rothe,Karsch:2002}. 
Actually, at the quenched level,
the deconfinement phase transition is of the 1st order, 
and both Polyakov loop $\langle L_P \rangle$ and 
the chiral condensate $\langle \bar qq \rangle$ 
jump at the same critical temperature $T_c$ \cite{Rothe}. 
(Note that the chiral condensate and the hadron masses 
can be calculated from the quark propagator 
even at the quenched level \cite{Rothe}.)

In the presence of dynamical quarks, 
the thermal phase transition of QCD is modified 
depending on the quark mass.
In the chiral limit of $N_f=3$, 
chiral transition is of the 1st order \cite{Rothe,Karsch:2002}.
For the two light u,d-quarks and relatively heavy s-quark
of $N_f = 2+1$, lattice QCD shows crossover at finite temperature, 
and hence, to be strict, there is no definite critical temperature
\cite{Karsch:2002,YAoki,AFKS06,AFKS09}.
Also in this case, almost coincidence of 
the two peak positions of the Polyakov-loop susceptibility 
and the chiral susceptibility suggests a close relation 
between confinement and chiral symmetry breaking in QCD \cite{Karsch:2002}, 
although several lattice-QCD studies reported a little higher 
peak position of the Polyakov-loop susceptibility than 
that of the light-quark chiral susceptibility \cite{AFKS06,AFKS09}.
In the case of QCD-like theory with adjoint-representation fermions, however, 
two phase transitions of deconfinement and chiral restoration 
occur at two distinct temperatures \cite{Karsch:1999,Engels:2005,Cossu}.

In the dual-superconductor picture \cite{DualSuperconductor}, 
the confinement is discussed in terms of the magnetic monopole 
which appears as the topological object in the maximally Abelian (MA) gauge 
\cite{Miyamura:1995,Woloshyn:1995, tHooft:1981,Kronfeld:1987,Hioki:1991,Stack:1994,
Suganuma:1995,Amemiya:1999,Gongyo:2013}. 
By removing magnetic monopoles from the QCD vacuum, 
both confinement and chiral symmetry breaking are simultaneously lost, 
as shown in lattice QCD \cite{Miyamura:1995, Woloshyn:1995}.
This fact suggests that both phenomena are related via magnetic monopoles. 
Similar results are also obtained 
by removing center vortices from the QCD vacuum 
in the maximal center gauge in lattice QCD
\cite{Gattner:2005, Hollwieser:2008}.
However, it is not sufficient to prove the direct relationship, 
since removing such topological objects might be too fatal 
for most non-perturbative QCD phenomena \cite{Suganuma:2011}.

On the other hand, 
the Dirac operator is directly related to chiral symmetry breaking.
As shown in the Banks-Casher relation, 
the chiral condensate $\langle \bar{q}q \rangle$ is 
proportional to the Dirac zero-mode density \cite{BanksCasher}, 
and chiral symmetry restoration is observed 
as a spectral gap of eigenmodes. 
Therefore, in order to clarify correspondence 
between chiral symmetry breaking and confinement, 
it is a promising approach to investigate 
confinement in terms of the Dirac eigenmodes 
\cite{Gattringer:2006,Bilgici:2008a,Bilgici:2008b, 
  Bruckmann:2007,Synatschke:2008,Hollwieser:2008,Kovacs:2010,
Lang:2011,Glozman:2012,Suganuma:2011,Gongyo:2012,Iritani:2012,TMDoi}.

In Gattringer's formula \cite{Gattringer:2006}, 
the Polyakov loop can be expressed by 
the sum of Dirac spectra with twisted boundary condition on lattice 
\cite{Gattringer:2006,Bruckmann:2007,Bilgici:2008a,
Bilgici:2008b,Synatschke:2008}.
In our previous studies, we formulated a gauge-invariant 
Dirac-mode expansion method in lattice QCD, 
and analyzed the contribution of Dirac-modes 
to the Wilson loop, the interquark potential, and the Polyakov loop 
\cite{Suganuma:2011,Gongyo:2012,Iritani:2012,TMDoi}.
In contrast to chiral symmetry breaking, 
these studies indicate that the low-lying Dirac eigenmodes 
are not relevant for confinement properties 
such as the area law of the Wilson loop, 
the linear confining potential, 
and the zero expectation value of the Polyakov loop 
\cite{Suganuma:2011,Gongyo:2012,Iritani:2012,TMDoi}.
It is also reported that hadrons still remain 
as bound states even without chiral symmetry breaking 
by removing low-lying Dirac-modes \cite{Lang:2011,Glozman:2012}.

In this paper, we perform the detailed analysis 
of the Polyakov loop in terms of Dirac eigenmodes, 
using the gauge-invariant Dirac-mode expansion method 
\cite{Suganuma:2011,Gongyo:2012}. 
In fact, we remove low-lying or high Dirac-modes 
from the QCD vacuum generated by lattice QCD, 
and then calculate the IR/UV-cut Polyakov loop 
in both confined and deconfined phases 
to investigate the contribution of the removed Dirac-modes 
to the confinement. 
We also discuss the temperature dependence of 
the IR/UV-cut Polyakov loop. 
For the Polyakov loop, unlike the Wilson loop, 
we can develop a practical calculation after removing 
the low-lying Dirac modes, by a reformulation 
with respect to the removed IR Dirac-mode space, 
which enables us to calculate with larger lattices. 

The organization of this paper is as follows.
In Sec.II, we briefly review the Dirac-mode expansion method 
in lattice QCD, and formulate the Dirac-mode projected Polyakov loop.
In Sec.III, we show the lattice QCD results of 
the Dirac-mode projected Polyakov loop 
in both confined and deconfined phases at finite temperature. 
In Sec.IV, we propose a new method to calculate the Polyakov loop 
without IR Dirac modes in a larger volume at finite temperature, 
by the reformulation with respect to the removed IR Dirac-mode space.
Section V will be devoted to summary.

\section{Formalism}
In this section, we review 
the Dirac-mode expansion method in lattice QCD 
\cite{Suganuma:2011,Gongyo:2012,Iritani:2012,TMDoi}, 
which is a gauge-invariant expansion 
with the Dirac eigenmode. 
We also formulate the Polyakov loop 
in the operator formalism of lattice QCD, 
and the Dirac-mode projected Polyakov loop. 
\cite{Gongyo:2012,Iritani:2012}.

\subsection{Dirac-mode expansion method in lattice QCD}

First, we briefly review the gauge-invariant formalism of 
the Dirac-mode expansion method in Euclidean lattice QCD 
\cite{Suganuma:2011,Gongyo:2012}.
The lattice-QCD gauge action is constructed 
from the link-variable $U_\mu(x) \in \mathrm{SU}(N_c)$, 
which is defined as $U_\mu(x) = e^{iagA_\mu(x)}$ 
with the gluon field $A_\mu(x) \in \mathfrak{su}(N_c)$, 
lattice spacing $a$, and gauge coupling constant $g$ \cite{Rothe}.
Using the link-variable $U_\mu(x)$, 
the Dirac operator $\Slash{D} = \gamma_\mu D_\mu$ is expressed as
\begin{equation}
  \Slash{D}_{x,y} \equiv \frac{1}{2a}
  \sum_{\mu = 1}^4
  \gamma_\mu \left[ U_\mu(x) \delta_{x+\hat{\mu},y}
  - U_{-\mu}(x) \delta_{x-\hat{\mu},y}\right]
  \label{eq:OriginalDirac}
\end{equation}
on lattice.  Here, we use the convenient notation of
$U_{-\mu}(x) \equiv U_\mu^\dagger(x-\hat{\mu})$,
and $\hat{\mu}$ denotes for the unit vector 
in $\mu$-direction in the lattice unit.
In this paper, the $\gamma$-matrix is defined to be hermitian,
i.e., $\gamma_\mu^\dagger = \gamma_\mu$.
Thus, the Dirac operator becomes an anti-hermitian operator as
\begin{equation}
  \Slash{D}_{y,x}^\dagger = - \Slash{D}_{x,y},
\end{equation}
and its eigenvalues are pure imaginary.
We introduce the normalized eigenstate $| n \rangle$,
which satisfies
\begin{equation}
  \Slash{D} | n \rangle = i \lambda_n | n \rangle
  \quad  (\lambda_n \in \mathbf{R})
\end{equation}
and $\langle n |m \rangle =\delta_{nm}$.
From the relation $\left\{ \gamma_5, \Slash{D} \right\} = 0$,
the eigenvalue appears as a pair $\{ i\lambda_n, - i\lambda_n \}$
for non-zero modes, 
since $\gamma_5 |n \rangle$ satisfies
$\Slash{D}\gamma_5 |n \rangle = - i \lambda_n \gamma_5 | n \rangle.$
The Dirac eigenfunction $\psi_n(x)$ is defined by 
\begin{equation}
  \psi_n(x) \equiv \langle x | n \rangle,
  \label{eq:DiracEigenFunction}
\end{equation}
and satisfies 
\begin{equation}
  \Slash{D}_{x,y}\psi_n(y) = i \lambda_n \psi_n(x).
\end{equation}
Considering the gauge transformation of the link-variable as
\begin{equation}
  U_\mu(x) \rightarrow V(x)U_\mu(x)V^\dagger(x+\mu)
\end{equation}
with SU($N_c$) matrix $V(x)$, 
the Dirac eigenfunction $\psi_n(x)$ is 
gauge-transformed like the matter field as \cite{Gongyo:2012}
\begin{equation}
  \psi_n(x) \rightarrow V(x) \psi_n(x).
  \label{eq:DiracEigenfunction}
\end{equation}
To be strict, in the transformation (\ref{eq:DiracEigenfunction}), 
there can appear an $n$-dependent irrelevant 
global phase factor $e^{i\phi_n}$, 
which originates from the arbitrariness of the definition of 
the eigenfunction $\psi_n(x)$ \cite{Gongyo:2012}.
However, such phase factors cancel between $|n\rangle$ and $\langle n|$, 
and do not appear in any gauge-invariant quantities such as 
the Wilson loop and the Polyakov loop. 

Next, we consider the operator formalism in lattice QCD, 
to keep the gauge covariance manifestly. 
We introduce the link-variable operator $\hat{U}_\mu$ 
defined by the matrix element as
\begin{equation}
  \langle x | \hat{U}_\mu | y \rangle = U_{\mu} (x) \delta_{x+\hat{\mu},y}.
\end{equation}
As the product of the link-variable operator $\hat{U}_\mu$, 
the Wilson loop operator $\hat{W}$ and the Polyakov loop operator $\hat{L}_P$ 
can be defined, 
and their functional trace, ${\rm Tr} \hat W$ and ${\rm Tr} \hat L_P$, 
are found to coincide with the Wilson loop $\langle W \rangle$ 
and the Polyakov loop $\langle L_P \rangle$, 
apart from an irrelevant constant factor \cite{Gongyo:2012}.
The Dirac-mode matrix element $\langle n | \hat{U}_\mu | m \rangle$ 
of the link-variable operator can be expressed as
\begin{eqnarray}
  \langle n | \hat{U}_\mu | m \rangle
  = \sum_x \langle n | x \rangle \langle x |
  \hat{U}_\mu | x + \hat{\mu} \rangle
  \langle x + \hat{\mu} | m \rangle 
  = \sum_x \psi_n^\dagger(x) U_\mu(x) \psi_m(x+\hat{\mu}),
\end{eqnarray}
by inserting $\sum_x |x\rangle \langle x|=1$ and 
using the Dirac eigenfunction (\ref{eq:DiracEigenFunction}).
The matrix element $\langle n | \hat{U}_\mu | m \rangle$
is gauge-transformed as
\begin{eqnarray}
  \langle n | \hat{U}_\mu | m \rangle
  &\rightarrow& \sum_x \psi_n^\dagger(x) V^\dagger(x) \cdot 
  V(x) U_\mu(x) V^\dagger(x+\hat{\mu}) \cdot V(x+\hat{\mu}) \psi_m(x+\hat{\mu}) \nonumber \\
  &=& \sum_x \psi_n^\dagger(x) U_\mu(x) \psi_m(x+\hat{\mu}) 
  = \langle n | \hat{U}_\mu | m \rangle.
\end{eqnarray}
Therefore, the matrix element 
$\langle n | \hat{U}_\mu | m \rangle$ is constructed in 
a gauge-invariant manner, apart from an irrelevant global phase factor
$e^{i\phi_n}$ \cite{Gongyo:2012}.

By inserting the completeness relation
\begin{eqnarray}
  \sum_n | n \rangle \langle n | = 1,
\end{eqnarray}
we can expand any operator $\hat{O}$ in terms of 
the Dirac-mode basis $|n\rangle$ as
\begin{equation}
  \hat{O} = \sum_n \sum_m | n \rangle \langle n | 
  \hat{O} | m\rangle\langle m|.
\end{equation}
This procedure is just an insertion of unity,
and it is mathematically correct.
This expression is the basis of the Dirac-mode expansion method.

Now, we introduce the Dirac-mode projection operator $\hat{P}$ as 
\begin{equation}
  \hat{P} \equiv \sum_{n \in \mathcal{A}} | n \rangle \langle n |
  \label{eq:DiracmodeProjection}
\end{equation}
for arbitrary subset $\mathcal{A}$ of the eigenmode space. 
For example, IR and UV mode-cut operators are given by
\begin{eqnarray}
  \hat{P}_{\rm IR} &\equiv&
  \sum_{|\lambda_n|\geq \Lambda_{\rm IR}}| n \rangle \langle n |, \\
  \hat{P}_{\rm UV} &\equiv& 
  \sum_{|\lambda_n|\leq \Lambda_{\rm UV}}| n \rangle \langle n |,
  \label{eq:IR-UV-cut-operator}
\end{eqnarray}
with the IR/UV cut $\Lambda_{\rm IR}$ and $\Lambda_{\rm UV}$.
Note that $\hat P$ satisfies $\hat P^2=\hat P$,
because of $\langle n|m\rangle=\delta_{nm}$.

Using the eigenmode projection operator,
we define Dirac-mode projected link-variable operator as
\begin{equation}
  \hat{U}_\mu^P \equiv \hat{P} \hat{U}_\mu \hat{P} 
  = \sum_{n \in \mathcal{A}} \sum_{m \in \mathcal{A}} 
  | n \rangle \langle n | \hat{U}_\mu | m \rangle \langle m |.
\end{equation}
By using this projected operator $\hat{U}_\mu^P$ 
instead of the original link-variable operator $\hat{U}_\mu$, 
we can analyze the contribution 
of individual Dirac eigenmode to the various quantities of QCD, 
such as the Wilson loop \cite{Suganuma:2011,Gongyo:2012}. 
  In general, this projection produces some non-locality.
  However, this non-locality would not be significant
  for the long-distance properties such as confinement
  \cite{Gongyo:2012}.

Here, we take the similar philosophy to clarify the importance of monopoles 
by removing them from the QCD vacuum 
\cite{Miyamura:1995,Woloshyn:1995,Stack:1994,Suganuma:1995}.
So far, by removing the monopoles from the gauge configuration 
generated by lattice QCD in the MA gauge and by checking its effect, 
several studies have shown 
the important role of monopoles to the nonperturbative phenomena such as 
confinement \cite{Rothe,Stack:1994}, 
chiral symmetry breaking \cite{Miyamura:1995,Woloshyn:1995}, 
and instantons \cite{Suganuma:1995}.

Note that, instead of the Dirac-mode basis, 
one can expand the link-variable operator 
with arbitrary eigenmode basis of appropriate operator in QCD.
For example, it would be also interesting to analyze the QCD phenomena 
in terms of the eigenmodes of 
the covariant Laplacian operator $D^2 = D^\mu D^\mu$ \cite{Bruckmann:2005}
and the Faddeev-Popov operator $M = - \partial_i D_i$ in the Coulomb gauge 
\cite{Gribov:1978,Zwanziger:2003,Greensite:2003,Greensite:2004,Greensite:2005}.

The advantages of the use of the Dirac operator 
are the gauge covariance and the Lorentz covariance.
In addition to these symmetries,
the Dirac operator is directly related 
to chiral symmetry breaking \cite{BanksCasher},
and also topological charge via Atiyah-Singer's 
index theorem \cite{AtiyahSinger}.
In fact, the chiral condensate $\langle \bar{q} q \rangle$ 
is proportional to the Dirac zero-mode density as 
\begin{equation}
  \langle \bar{q}q \rangle  = - \lim_{m \rightarrow 0}
  \lim_{V\rightarrow \infty} \pi \rho(0),
\end{equation}
which is known as the Banks-Casher relation \cite{BanksCasher}. 
Here, $\rho(\lambda)$ is the spectral density of the Dirac operator, 
and is given by
\begin{equation}
  \rho(\lambda) \equiv \frac{1}{V_{\rm phys}} \sum_n
  \langle \delta(\lambda-\lambda_n) \rangle,
\end{equation}
with four-dimensional space-time volume $V_{\rm phys}$.

We also note that the low-lying Dirac-mode is closely related 
to instantons. 
By filtering ultraviolet eigenmodes, 
instanton-like structure is clearly revealed
without cooling or smearing techniques \cite{Ilgenfritz:2007}.
In fact, Dirac eigenfunctions are useful probes to 
investigate the topological structure of the QCD vacuum.

For the Dirac-mode expansion, 
we use the lattice Dirac operator \eqref{eq:OriginalDirac}. 
To reduce the computational cost, 
we utilize the Kogut-Susskind (KS) formalism \cite{Rothe}, 
and deal with the KS Dirac operator, 
\begin{equation}
  D^{\rm KS}_{x,y} \equiv \frac{1}{2a}
  \sum_{\mu = 1}^{4} \eta_\mu(x) 
  \left[ U_\mu(x) \delta_{x+\mu,y}
  - U_{-\mu}(x) \delta_{x-\mu,y} \right],
\end{equation}
with the staggered phase $\eta_\mu(x)$ defined by
\begin{equation}
  \eta_1(x) \equiv 1, \quad
  \eta_\mu(x) \equiv (-1)^{x_1 + \cdots + x_{\mu-1}}
  \quad (\mu \geq 2).
\end{equation}
Using KS operator basis, one can drop off the spinor index, 
and it reduces the computational cost.
For the calculation of the Polyakov loop, it can be proven that 
the KS Dirac-mode expansion gives the same result 
as the original Dirac-mode expansion \cite{TMDoi}.

\subsection{Polyakov-loop operator and its Dirac-mode projection}
Next, we formulate the Polyakov loop in the operator formalism, 
and the Dirac-mode projected Polyakov loop in SU(3) lattice QCD 
with the space-time volume $V = L^3 \times N_t$ and 
the ordinary periodic boundary condition.
Using the temporal link-variable operator $\hat{U}_4$, 
the Polyakov-loop operator $\hat{L}_P$ is defined as 
\begin{equation}
  \hat{L}_P \equiv \frac{1}{3V} \prod_{i=1}^{N_t} \hat{U}_4
  = \frac{1}{3V} \hat{U}_4^{N_t}
\end{equation}
in the operator formalism.
By taking the functional trace ``Tr'',
the Polyakov-loop operator leads to 
the expectation value of the 
ordinary Polyakov loop $\langle L_P \rangle$ as
  \begin{eqnarray}
    \mathrm{Tr} \ \hat{L}_P 
    &=& \frac{1}{3V} \mathrm{Tr} \ \{ \prod_{i=1}^{N_t} \hat{U}_4 \} 
    = \frac{1}{3V} \mathrm{tr}
    \sum_{\vec{x},t} \langle \vec{x},t | \prod_{i=1}^{N_t} \hat{U}_4 | \vec{x},t \rangle 
    \nonumber  \\
    &=& \frac{1}{3V} \mathrm{tr} \sum_{\vec{x},t} 
    \langle \vec{x},t | \hat{U}_4 | \vec{x},t + a \rangle 
    \langle \vec{x},t+a | \hat{U}_4 | \vec{x},t+2a \rangle 
    \cdots
    \langle \vec{x},t + (N_t-1)a | \hat{U}_4 | \vec{x},t \rangle \nonumber \\
    &=& \frac{1}{3V} \mathrm{tr} \sum_{\vec{x},t} U_4(\vec{x},t) U_4(\vec{x},t+a) 
    \cdots U_4(\vec{x},t+(N_t-1)a) 
    = \langle L_P \rangle
  \end{eqnarray}
In this paper, we use``tr'' for the trace over SU(3) color index.
Using the Dirac-mode projection operator $\hat{P}$ 
in Eq.~(\ref{eq:DiracmodeProjection}),
we define Dirac-mode projected Polyakov-loop operator 
$L_P^{\rm proj.}$ as \cite{Gongyo:2012, Iritani:2012}
  \begin{eqnarray}
    {L}_P^{\rm proj.} &\equiv&
    \frac{1}{3V} \mathrm{Tr} \prod_{i=1}^{N_t} \{ \hat{U}_4^P \} 
    =\frac{1}{3V} \mathrm{Tr} \{ (\hat{U}_4^P)^{N_t} \} 
    =\frac{1}{3V} \mathrm{Tr} \{ (\hat U_4 \hat P)^{N_t} \} \nonumber \\
    &=& \frac{1}{3V}\mathrm{Tr}\{ \hat{P}\hat{U}_4\hat{P}\hat{U}_4\hat{P}
    \cdots \hat{P}\hat{U}_4\hat{P}\} \nonumber \\
    &=& \frac{1}{3V}
    \mathrm{tr} \sum_{n_1,n_2,\dots,n_{N_t} \in \mathcal{A}}
    \langle n_1 | \hat{U}_4 | n_2 \rangle
    \langle n_2 | \hat{U}_4 | n_3 \rangle 
    \cdots \langle n_{N_t} | \hat{U}_4 | n_1 \rangle.
    \label{eq:Lproj}
  \end{eqnarray}
Similar to Gattringer's formula \cite{Gattringer:2006}, 
we can investigate the contribution of the individual Dirac-mode 
to the Polyakov loop using this formula \eqref{eq:Lproj}.
In this paper, 
we mainly analyze the effect of removing low-lying (IR) and
high (UV) Dirac-modes, respectively, 
and denote IR/UV-mode cut Polyakov loop as
\begin{eqnarray}
  \langle L_P \rangle_{\rm IR} \equiv \frac{1}{3V}
  \mathrm{tr} \sum_{|\lambda_{n_i}| \geq \Lambda_{\rm IR}}
  \langle n_1 | \hat{U}_4 | n_2 \rangle \cdots
  \langle n_{N_t} | \hat{U}_4 | n_1 \rangle, \\
  \langle L_P \rangle_{\rm UV} \equiv \frac{1}{3V}
  \mathrm{tr} \sum_{|\lambda_{n_i}| \leq \Lambda_{\rm UV}}
  \langle n_1 | \hat{U}_4 | n_2 \rangle \cdots
  \langle n_{N_t} | \hat{U}_4 | n_1 \rangle,
\end{eqnarray}
with the IR/UV-cut parameter 
$\Lambda_{\rm IR}$/$\Lambda_{\rm UV}$.
We also investigate the effect of removing 
intermediate Dirac-modes in Appendix A.
  Note that,
  even the non-locality appears through the Dirac-mode projection,
  its effect just gives an extension to the Polyakov line,
  so that its infrared effect should be negligible 
  for the Polyakov loop \cite{Gongyo:2012}.

\section{Lattice QCD Results}
We study the Polyakov loop and the $Z_3$ center symmetry 
in terms of the Dirac mode in SU(3) lattice QCD at the quenched level, 
using the standard plaquette action and 
the ordinary periodic boundary condition.
We adopt the jackknife method to estimate the statistical error.

In this section, we calculate full eigenmodes of the Dirac operator 
using LAPACK \cite{LAPACK}, and perform the Dirac-mode removal 
from the nonperturbative vacuum generated by lattice QCD calculations.
We investigate the Polyakov loop without the specific Dirac-modes, 
showing the full figure of the Dirac spectrum.
For the reduction of the computational cost, 
we utilize the Kogut-Susskind (KS) formalism \cite{Gongyo:2012}. 
However, to obtain the full Dirac eigenmodes, 
the reduced computational cost is still quite large, 
and then we take relatively small lattices, 
$6^4$ and $6^3 \times 4$.
The calculation with larger lattice of $12^3 \times 4$ 
will be discussed in Sec.IV.

\subsection{Dirac-mode projected Polyakov loop in the confined phase}

In this subsection, 
we mainly analyze the contribution of the low-lying Dirac modes to 
the Polyakov loop in the confined phase below $T_c$.

Below $T_c$, the expectation value of the Polyakov loop 
$\langle L_P\rangle$ is very small, 
and would be exactly zero in infinite volume. 
Then, one may simply consider that any part of zero is zero 
and any type of filtering leaves zero unchanged.
However, this is not correct. 
For example, after the filtering of the monopole removal in the MA gauge,
the Polyakov loop has a non-zero expectation value 
in the remaining system called as the photon part, 
even at low temperatures \cite{Miyamura:1995}. 
Similarly, the confinement property is lost by 
the removal of center vortices in the maximal center gauge 
\cite{Debbio, deForcrand}, or by cutting off the infrared-momentum 
gluons in the Landau gauge \cite{Yamamoto}. 
We also comment on the other filtering of smearing and cooling methods, 
which are popular techniques to remove quantum fluctuations \cite{Rothe}.
Using these methods, the low-lying Dirac eigenmode density is reduced 
\cite{Gattringer:2006wq}, 
and confinement property will be lost after many iterations of the cooling.
These filtering operations actually change the Polyakov loop 
from zero even at low temperatures. 
In fact, after some filtering, it is nontrivial whether 
the system keeps $\langle L_P\rangle=0$ or not.

Here, we consider an interesting filtering of 
the Dirac-mode projection, introduced in the previous section.
We use the periodic $6^4$ lattice with $\beta = 5.6$ 
at the quenched level. 
The lattice spacing $a$ is found to be about $0.25$~fm 
\cite{Suganuma:2011,Gongyo:2012}, which is determined 
so as to reproduce the string tension 
$\sigma = 0.89$~GeV/fm \cite{Takahashi}.
If one regards this system as the finite temperature system, 
the temperature is estimated as 
$T = 1/(N_t a) \simeq 0.13$~GeV.

\begin{figure}
  \centering
  \includegraphics[width=0.50\textwidth,clip]{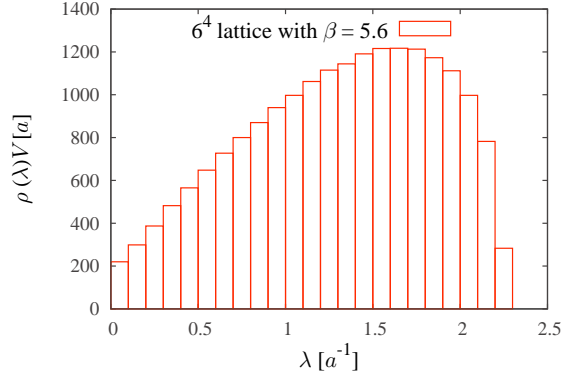}
  \caption{\label{fig:DiracSpectrum}
  The Dirac spectral density $\rho(\lambda)$
  on $6^4$ lattice with $\beta = 5.6$, i.e., $a \simeq 0.25$~fm.
  Because of $\rho(-\lambda)=\rho(\lambda)$, 
  only the positive region of $\lambda$ is shown.
  The bin-width is taken as $\Delta \lambda = 0.1a^{-1}$.
  The total number of eigenmodes is $6^4 \times 3 = 3888$.
}
  \end{figure}

We show the Dirac-spectral density $\rho(\lambda)$ 
in Fig.~\ref{fig:DiracSpectrum}.
The total number of eigenmodes is $6^4 \times 3 = 3888$. 
From this spectral density, we remove the low-lying or high eigenmodes, 
and analyze their contribution to the Polyakov loop, respectively. 
The Banks-Casher relation shows that 
the low-lying Dirac-modes are the essential ingredient for the chiral condensate
$\langle \bar{q} q \rangle$.
With the IR Dirac-mode cut $\Lambda_{\rm IR}$,
the chiral condensate is given by
\begin{equation}
  \langle \bar{q}q \rangle_{\rm IR}
  = - \frac{1}{V} \sum_{\lambda_n \geq \Lambda_{\rm IR}}
  \frac{2m}{\lambda_n^2 + m^2},
\end{equation}
where $m$ is the current quark mass.

\begin{figure}
    \centering
    \includegraphics[width=0.45\textwidth]{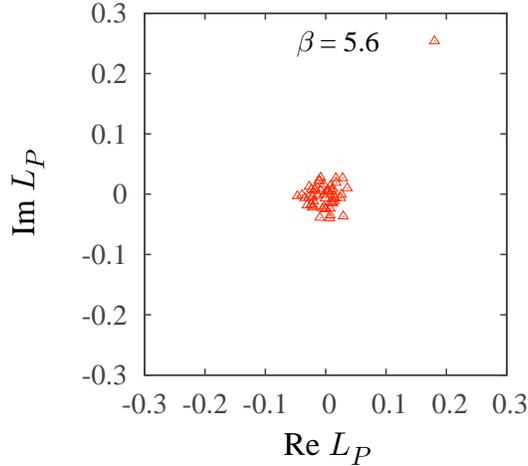}
    \caption{\label{fig:PolyakovScatterConfined} 
    The scatter plot of the Polyakov loop 
    $\langle L_P \rangle$ 
    in the confined phase 
    on the periodic lattice of $6^4$ and $\beta = 5.6$, i.e.,
    $a \simeq 0.25$~fm.
  }
\end{figure}

  Figure \ref{fig:PolyakovScatterConfined} is 
  the scatter plot of the original (no Dirac-mode cut) Polyakov loop 
  $\langle L_P \rangle$
  for 50 gauge configurations.
  As shown in Fig.~\ref{fig:PolyakovScatterConfined},
 $\langle L_P \rangle$ 
  is almost zero, 
  and $Z_3$-center symmetry is unbroken.

\begin{figure}
  \centering
  \includegraphics[width=0.50\textwidth]{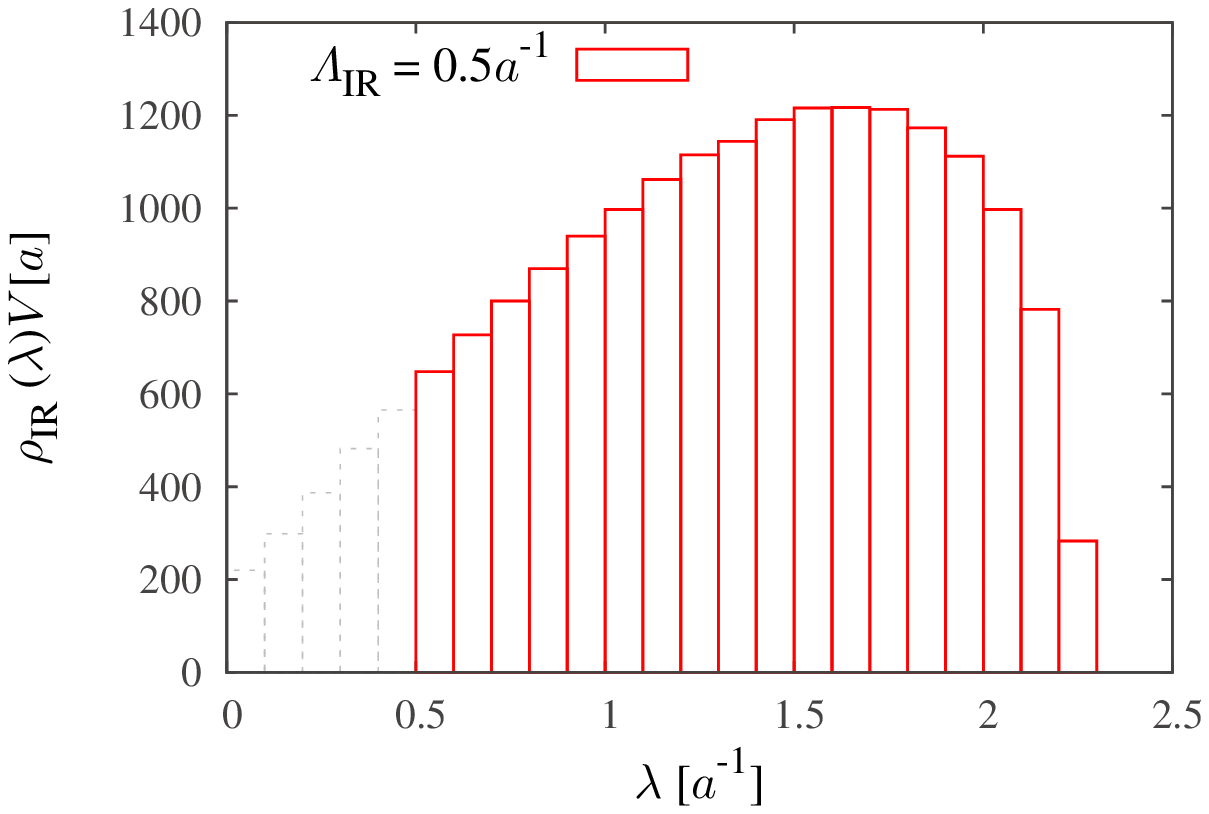}
  \includegraphics[width=0.45\textwidth]{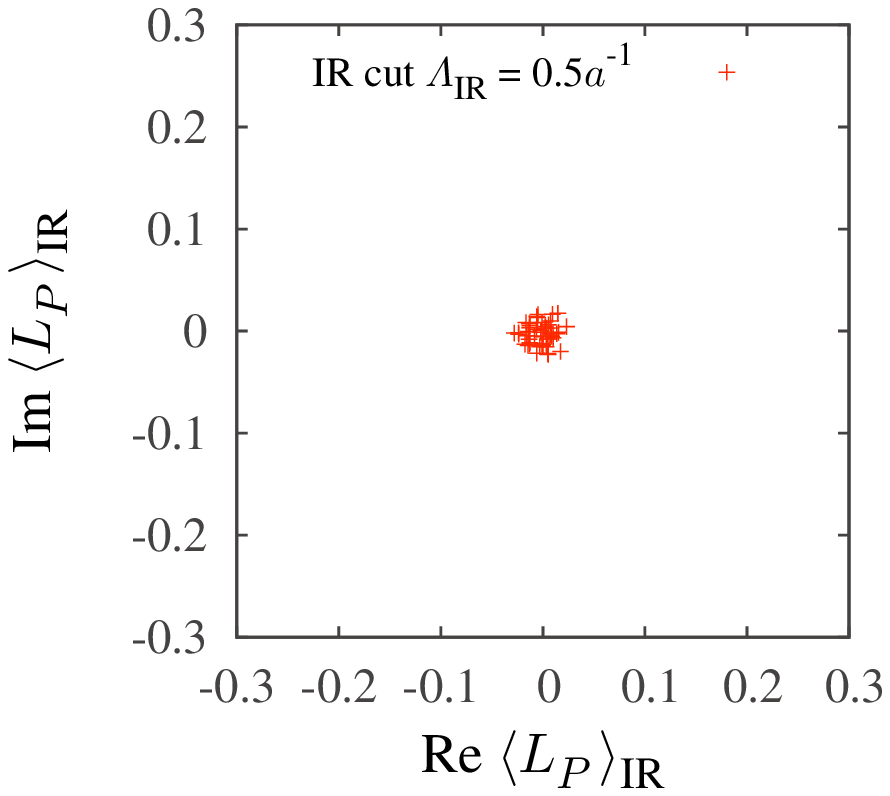}
  \caption{\label{fig:PolyakovScatterConfinedIRcut} 
  (a) The IR-cut Dirac spectral density 
  $\rho_{\rm IR}(\lambda)\equiv 
  \rho(\lambda)\theta(|\lambda|-\Lambda_{\rm IR})$ and 
  (b) the IR-cut Polyakov loop 
  $\langle L_P\rangle_{\rm IR}$ 
  on the periodic lattice of $6^4$ at $\beta = 5.6$ 
  for the IR-cut of $\Lambda_{\rm IR} = 0.5a^{-1}$.
}
  \end{figure}

First, we analyze the role of low-lying Dirac-modes 
using the 50 gauge configurations. 
Figure \ref{fig:PolyakovScatterConfinedIRcut} shows IR-cut spectral density 
\begin{equation}
  \rho_{\rm IR}(\lambda) \equiv \rho(\lambda)\theta(|\lambda|-\Lambda_{\rm IR}),
\end{equation}
and the scatter plot of 
the IR-cut Polyakov loop 
$\langle L_P \rangle_{\rm IR}$ 
for $\Lambda_{\rm IR} = 0.5a^{-1}$,
which corresponds to about 400 modes removing from full eigenmodes. 
By this removal of low-lying Dirac modes below 
$\Lambda_{\rm IR} = 0.5a^{-1} \simeq 0.4$~GeV, 
the IR-cut chiral condensate 
$\langle \bar{q}q \rangle_{\rm IR}$ is extremely reduced as 
\begin{equation}
  {\langle \bar{q}q \rangle_{\rm IR}}/
  {\langle \bar{q}q \rangle}\simeq 0.02
\end{equation}
around the physical region of the current quark mass, 
$m \simeq 0.006a^{-1}\simeq 5$~MeV \cite{Gongyo:2012}.

As shown in Fig.~\ref{fig:PolyakovScatterConfinedIRcut}(b), 
even without the low-lying Dirac-modes, 
the IR-cut Polyakov loop is still almost zero \cite{Gongyo:2012}, 
\begin{equation}
  \langle L_P \rangle_{\rm IR} \simeq 0  
\end{equation}
and $Z_3$-center symmetry is unbroken. 
This result shows that the single-quark energy remains extremely large, 
and the system is still in the confined phase 
even without low-lying Dirac-modes.

\begin{figure}
    \centering
    \includegraphics[width=0.50\textwidth]{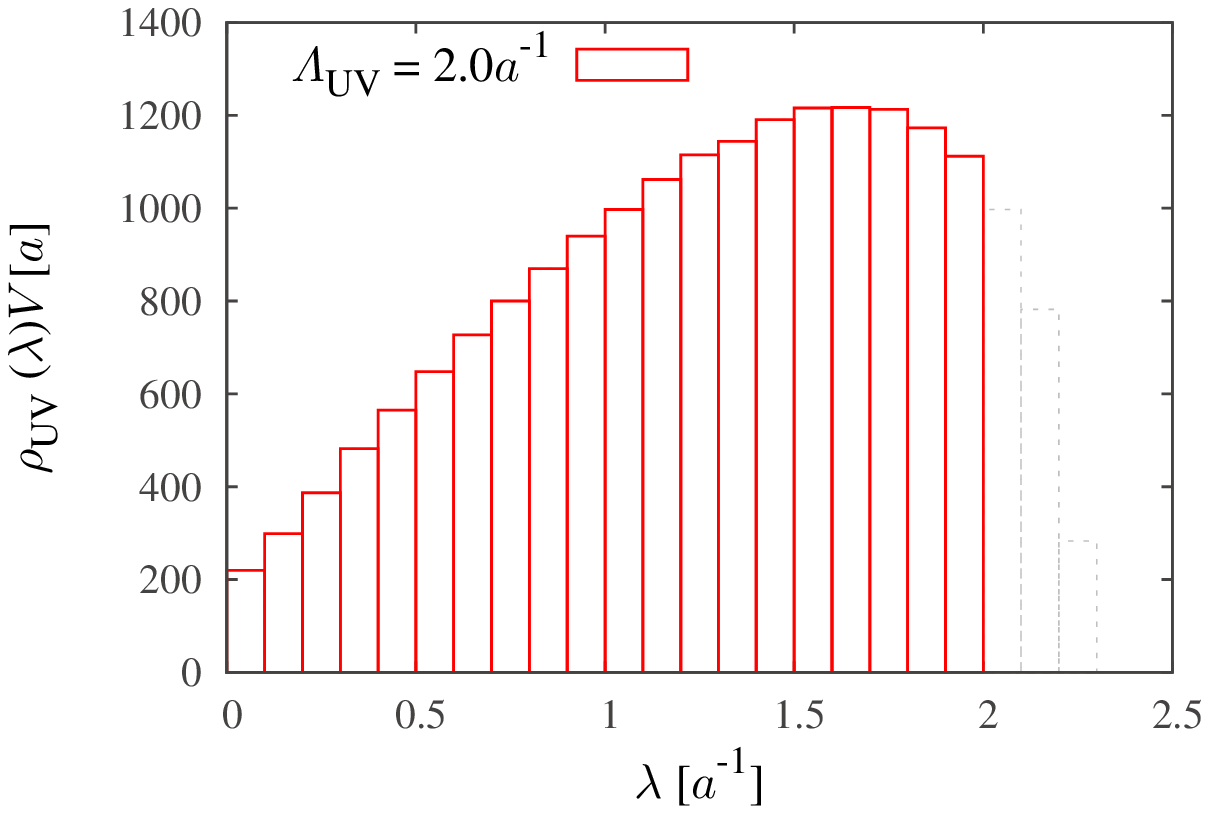}
    \includegraphics[width=0.45\textwidth]{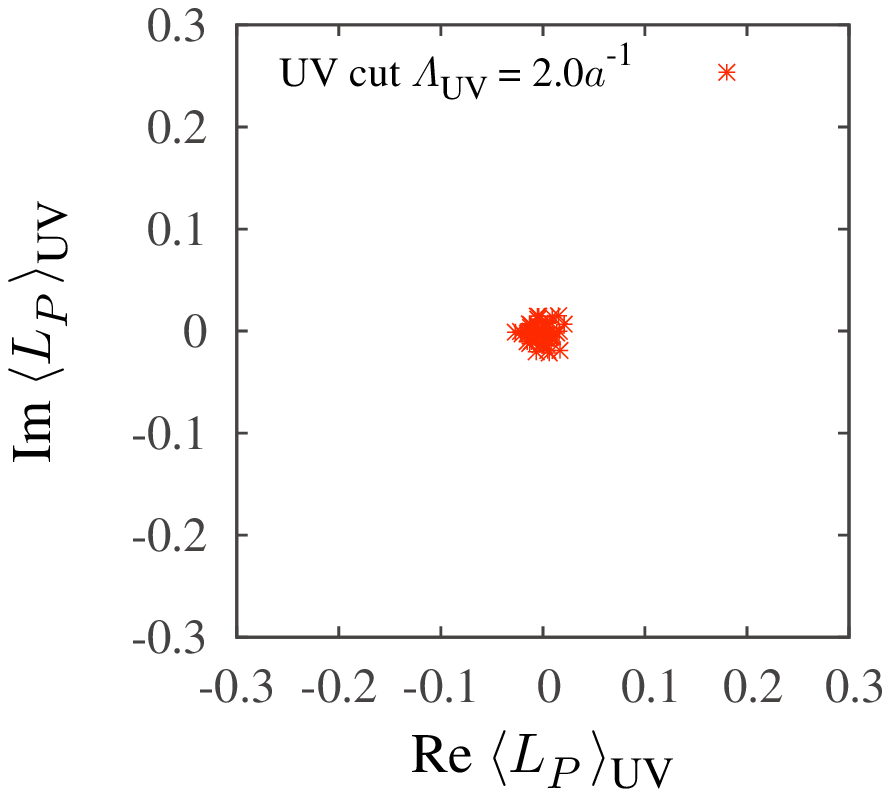}
    \caption{\label{fig:PolyakovScatterConfinedUV}
    (a) The UV-cut Dirac spectral density 
    $\rho_{\rm UV}(\lambda)\equiv 
    \rho(\lambda)\theta(\Lambda_{\rm UV}-|\lambda|)$ and 
    (b) the UV-cut Polyakov loop $\langle L_P\rangle_{\rm UV}$ 
    on the periodic lattice of $6^4$ at $\beta = 5.6$ 
    for $\Lambda_{\rm UV} = 2.0a^{-1}$.
  }
\end{figure}

  Second, we consider the high Dirac-mode contribution to the Polyakov loop 
  in the confined phase below $T_c$. 
  In this case, the chiral condensate is almost unchanged. 
  Figure \ref{fig:PolyakovScatterConfinedUV} shows the UV-cut spectral density 
  \begin{eqnarray}
    \rho_{\rm UV}(\lambda) \equiv \rho(\lambda)\theta(\Lambda_{\rm UV}-|\lambda|),
  \end{eqnarray}
  and the UV-cut Polyakov loop 
  $\langle L_P\rangle_{\rm UV}$ 
  for $\Lambda_{\rm UV} = 2.0a^{-1}$, 
  corresponding to the removal of about 400 modes. 
  Similar to the cut of low-lying modes, 
  the UV-cut Polyakov loop is almost zero 
  as 
  $\langle L_P\rangle_{\rm UV}\simeq 0$, 
  and indicates the confinement.

Thus, in both cuts of low-lying Dirac modes 
in Fig.~\ref{fig:PolyakovScatterConfinedIRcut}(b)
and high modes in Fig.~\ref{fig:PolyakovScatterConfinedUV}(b), 
the Polyakov loop 
$\langle L_P\rangle_{\rm IR/UV}$ is almost zero, which 
means that the system remains in the confined phase.
In fact, we find ``Dirac-mode insensitivity'' to 
the Polyakov loop or the confinement property. 
We also examine the removal of intermediate (IM) Dirac-modes 
from the Polyakov loop in the confined phase in Appendix A, 
and find the similar Dirac-mode insensitivity.
It suggests that 
each eigenmode has the information of confinement,
and the Polyakov loop is not affected by removing of any eigenvalue region.
Therefore, we consider that there is no direct correspondence 
between the Dirac eigenmodes and the Polyakov loop in the confined phase.
This Dirac-mode insensitivity to confinement is consistent 
with the previous Wilson-loop analysis \cite{Suganuma:2011,Gongyo:2012}.

\subsection{Dirac-mode projected Polyakov loop in the deconfined phase}
Next, we investigate the Polyakov loop 
in the deconfined phase at high temperature.
Here, we use periodic lattice of $6^3 \times 4$ at $\beta = 6.0$, 
which corresponds to 
$a \simeq 0.10$~fm and $T \equiv 1/(N_ta) \simeq 0.5$~GeV.

\begin{figure}
  \centering
  \includegraphics[width=0.45\textwidth]{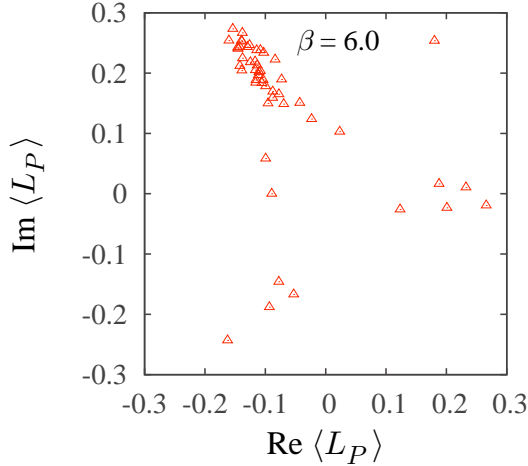}
  \caption{\label{fig:PolyakovScatterDeconfined}
  The scatter plot of the Polyakov loop $L_P$ 
  in the deconfined phase on the periodic lattice of 
  $6^3 \times 4$ at $\beta = 6.0$, corresponding to 
  $a \simeq 0.10$~fm and $T \equiv 1/(N_ta) \simeq 0.5$~GeV.
}
  \end{figure}

As shown in Fig.~\ref{fig:PolyakovScatterDeconfined}, 
the Polyakov loop has non-zero expectation values as 
$\langle L_P \rangle \neq 0$, and shows $Z_3$-center group structure 
on the complex plane. This behavior means the deconfined 
and center-symmetry broken phase.

\begin{figure}
    \centering
    \includegraphics[width=0.47\textwidth,clip]{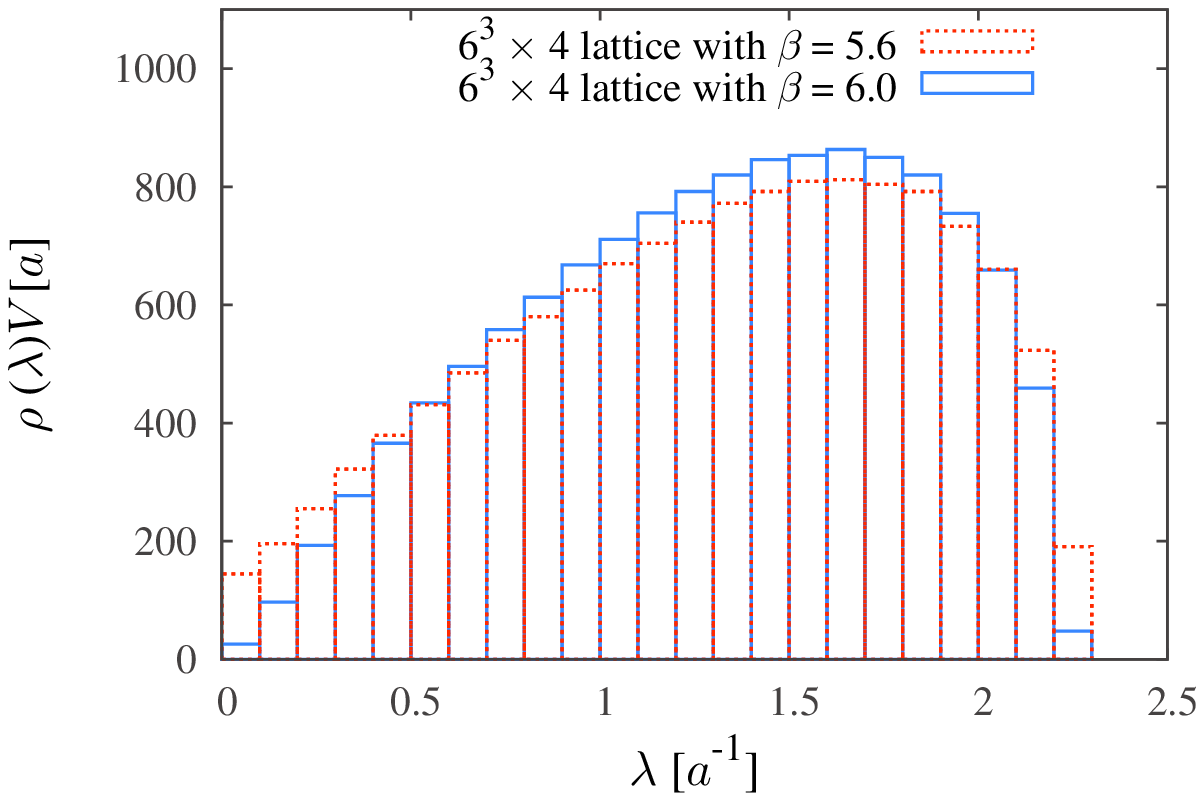}
    \includegraphics[width=0.47\textwidth,clip]{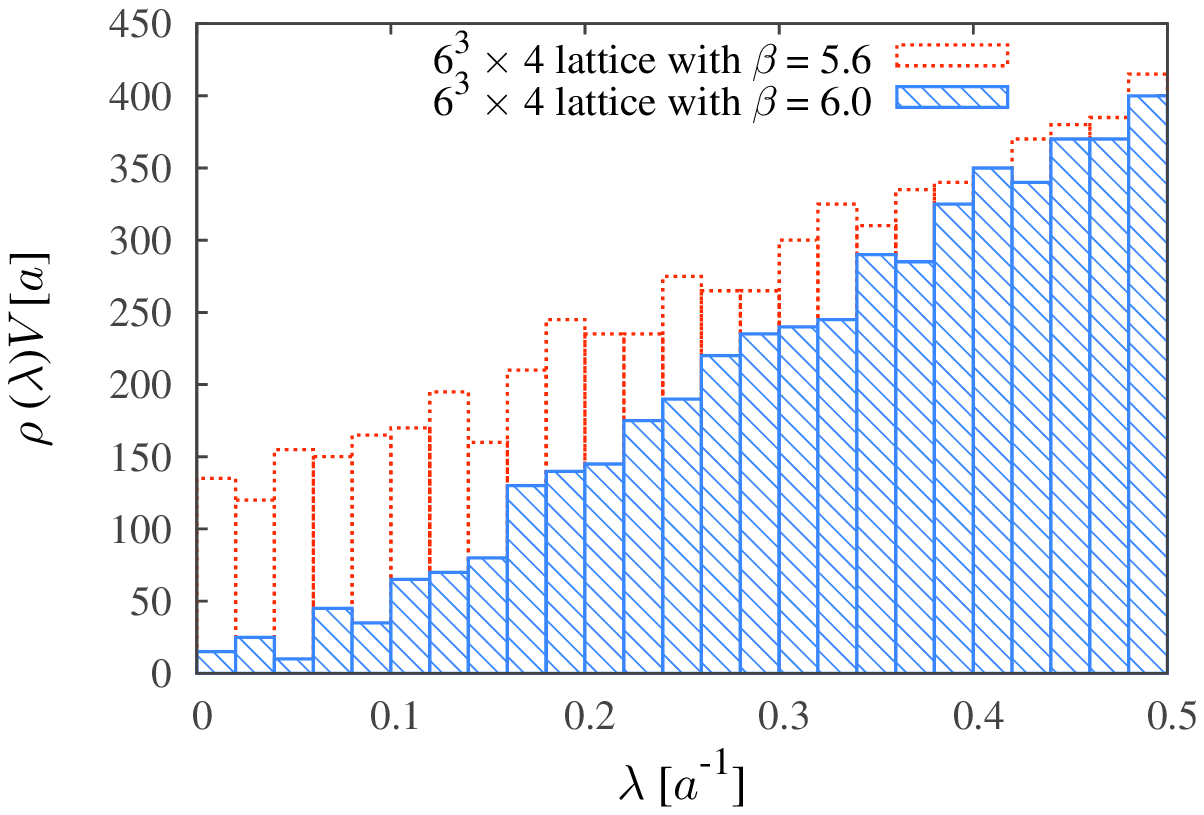}
    \caption{ \label{fig:DiracSpectrumDeconfined}
    The Dirac spectrum density $\rho(\lambda)$ 
    in confined phase ($\beta = 5.6$) 
    and deconfined phase ($\beta = 6.0$) 
    on $6^3 \times 4$ lattice. 
    (a) The comparison on full spectral densities. 
    (b) The comparison on low-lying spectral densities.
  }
\end{figure}

  To begin with, we investigate the difference of 
  the Dirac spectral density $\rho(\lambda)$
  between the confined and the deconfined phases.
  Figure~\ref{fig:DiracSpectrumDeconfined} shows 
  the Dirac spectral density in the deconfined phase 
  at high temperature on $6^3 \times 4$ at $\beta = 6.0$, i.e., 
  $T \simeq 0.5$~GeV. 
  For comparison, we also add the spectrum density 
  in the confined phase at low temperature on 
  $6^3 \times 4$ at $\beta = 5.6$, i.e., 
  $a \simeq 0.25$~fm and $T = 1/(N_ta) \simeq 0.2$~GeV, 
  below the critical temperature $T_c \simeq 0.26$~GeV 
  at the quenched level.
  In both phases, 
  the total number of eigenmodes is $6^3 \times 4 \times 3 = 2592$.
  As shown in Fig.~\ref{fig:DiracSpectrumDeconfined},
  the low-lying Dirac eigenmodes are suppressed in the 
  high-temperature phase, which leads to the chiral restoration.

\begin{figure}
  \centering
  \includegraphics[width=0.45\textwidth]{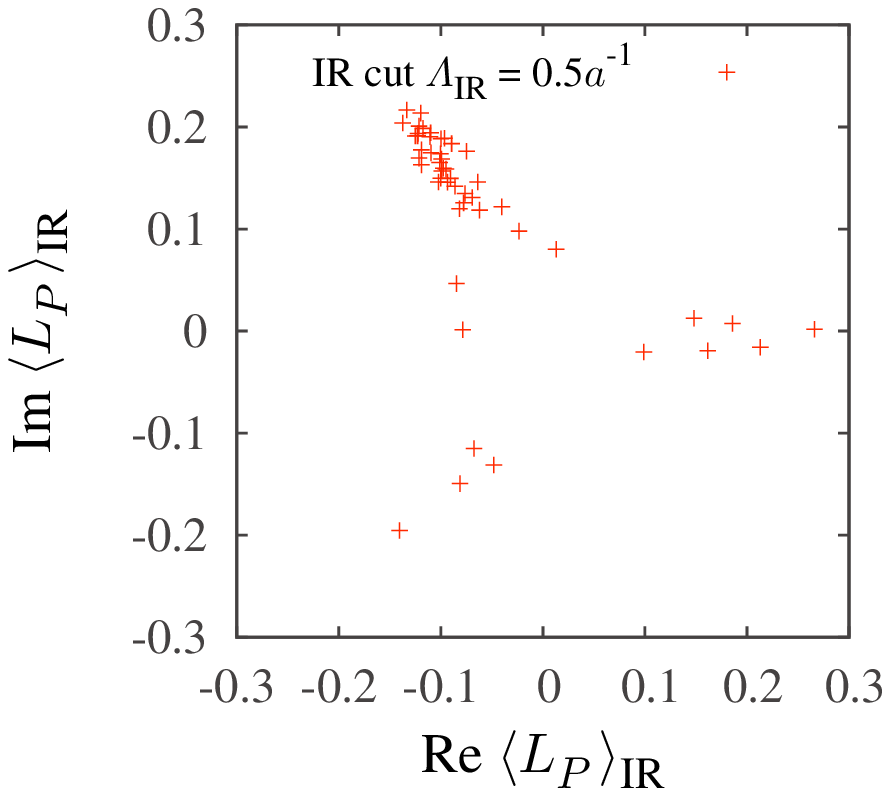}
  \includegraphics[width=0.45\textwidth]{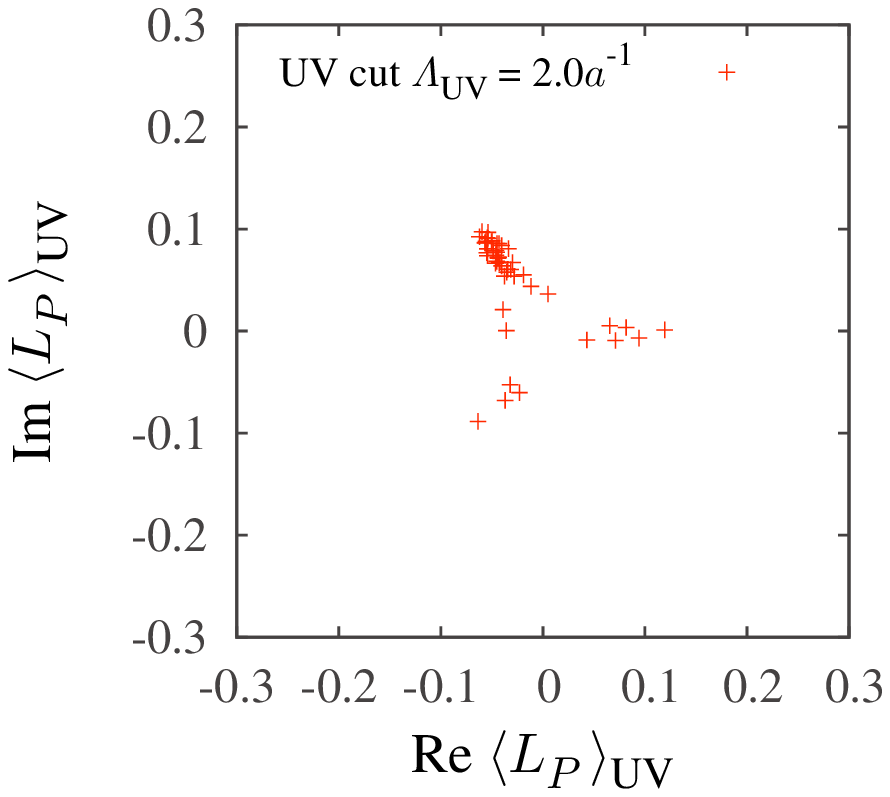}
  \caption{\label{fig:PolyakovScatterDeconfinedCut}
  The scatter plot of the IR/UV-cut Polyakov loop 
  in the deconfined phase on the periodic lattice of 
  $6^3 \times 4$ at $\beta = 6.0$, 
  i.e., $a \simeq 0.10$~fm and $T \equiv 1/(N_ta) \simeq 0.5$~GeV. 
  (a) $\langle L_P\rangle_{\rm IR}$ in the case of IR Dirac-mode cut of 
  $\Lambda_{\rm IR} = 0.5a^{-1}$. 
  (b) $\langle L_P\rangle_{\rm UV}$ in the case of UV Dirac-mode cut of 
  $\Lambda_{\rm UV} = 2.0a^{-1}$. 
  According to the mode cut, there appears a constant reduction factor.
}
\end{figure}

We show the Dirac-mode projected Polyakov loop 
$\langle L_P\rangle_{\rm IR/UV}$ 
at $\Lambda_{\rm IR} = 0.5a^{-1}$ and $\Lambda_{\rm UV} = 2.0a^{-1}$ 
in Figs.~\ref{fig:PolyakovScatterDeconfinedCut}(a) and (b), respectively.
These mode-cuts correspond to removing about 200 modes from full eigenmodes.
According to the removal of about 200 modes, there appears 
a trivial reduction (or normalization) factor for 
the IR/UV-cut Polyakov loop $\langle L_P\rangle_{\rm IR/UV}$.

As shown in Fig.~\ref{fig:PolyakovScatterDeconfinedCut}, 
both IR/UV-cut Polyakov loops $\langle L_P\rangle_{\rm IR/UV}$ 
are non-zero and show the characteristic $Z_3$ structure, 
similar to the original Polyakov loop $\langle L_P \rangle$. 
This suggests Dirac-mode insensitivity also in the deconfined phase.
In Appendix A, 
we show the IM-cut Polyakov loop $\langle L_P\rangle_{\rm IM}$ 
in the deconfined phase, and find the similar results.

We also note that the absolute value of UV-cut Polyakov loop 
$\langle L_P\rangle_{\rm UV}$ is smaller than that of 
IR-cut one $\langle L_P\rangle_{\rm IR}$ in each gauge configuration, 
as shown in Fig.~\ref{fig:PolyakovScatterDeconfinedCut}, 
in spite of almost the same number of removed IR/UV-modes. 
In fact, as the quantitative effect to the Polyakov loop, 
the contribution of UV Dirac-modes 
is larger than that of IR Dirac-modes \cite{Bruckmann:2007},
although the deconfinement nature indicated by 
the non-zero Polyakov loop 
does not change by the removal of IR or UV Dirac-modes.

Thus, the Polyakov-loop behavior and the $Z_3$ center symmetry 
are rather insensitive to the removal of the Dirac-modes 
in the IR, IM or UV region in both confined and deconfined phases. 
Therefore, we conclude that there is no clear correspondence 
between the Dirac-modes and the Polyakov loop 
in both confined and deconfined phases.

\subsection{Temperature dependence of the Dirac-mode projected Polyakov loop}

So far, we have analyzed the role of the Dirac-mode to 
the Polyakov loop in both confined and deconfined phases.
In this subsection, we consider the temperature dependence 
of the Polyakov loop in terms of the Dirac-mode 
by varying the lattice parameter $\beta$ at fixed $N_t$.
Here, we use $6^3 \times 4$ lattice with $\beta = 5.4 \sim 6.0$.

Now, we investigate the gauge-configuration average of
the absolute value of the IR/UV-cut Polyakov loop,
\begin{eqnarray}
\langle |L_P^{\rm IR/UV}| \rangle
\equiv \frac{1}{N_{\rm conf}}
\sum_{k=1}^{N_{\rm conf}}|\langle L_P \rangle^{\rm IR/UV}_{k}|,
\end{eqnarray}
where 
$\langle L_P \rangle^{\rm IR/UV}_k$ denotes 
the IR/UV-cut Polyakov loop obtained from $k$-th gauge configuration, 
and $N_{\rm conf}$ the gauge configuration number.

\begin{figure}
  \begin{center}
    \includegraphics[width=0.48\textwidth,clip]{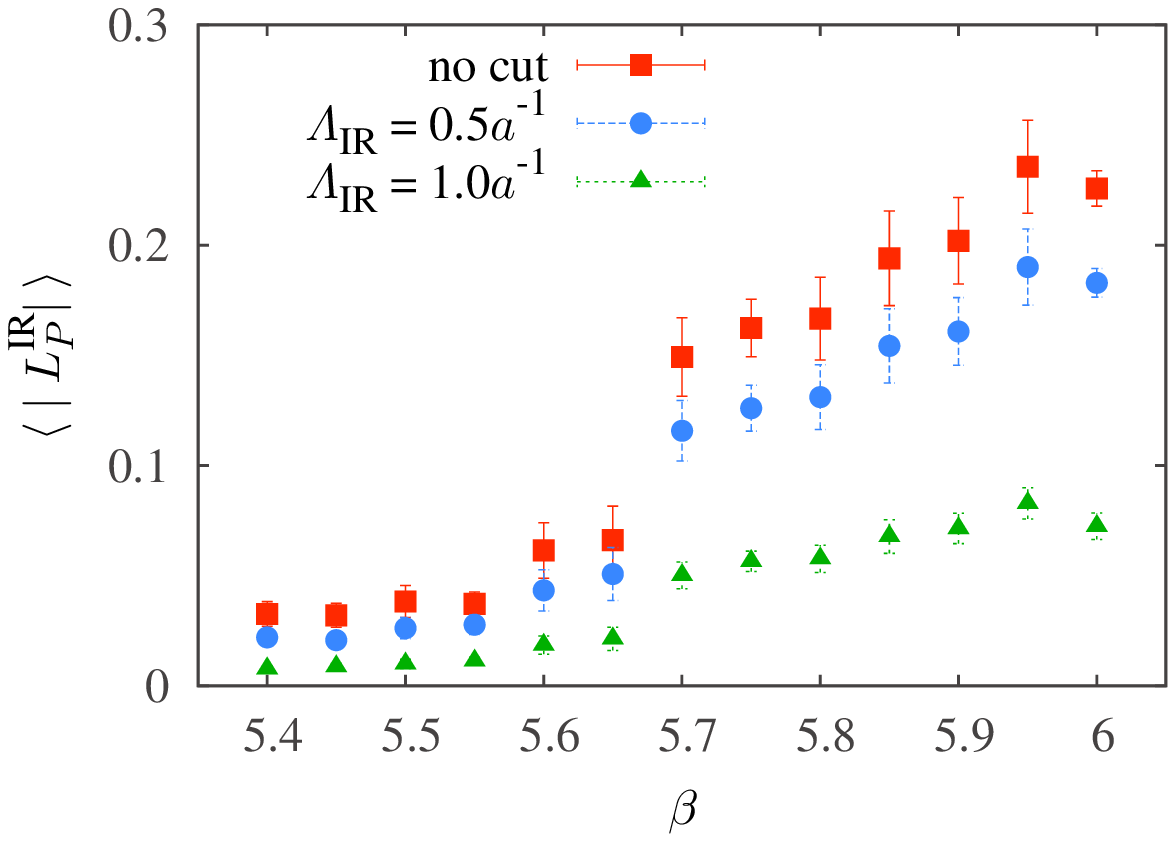}
    \includegraphics[width=0.48\textwidth,clip]{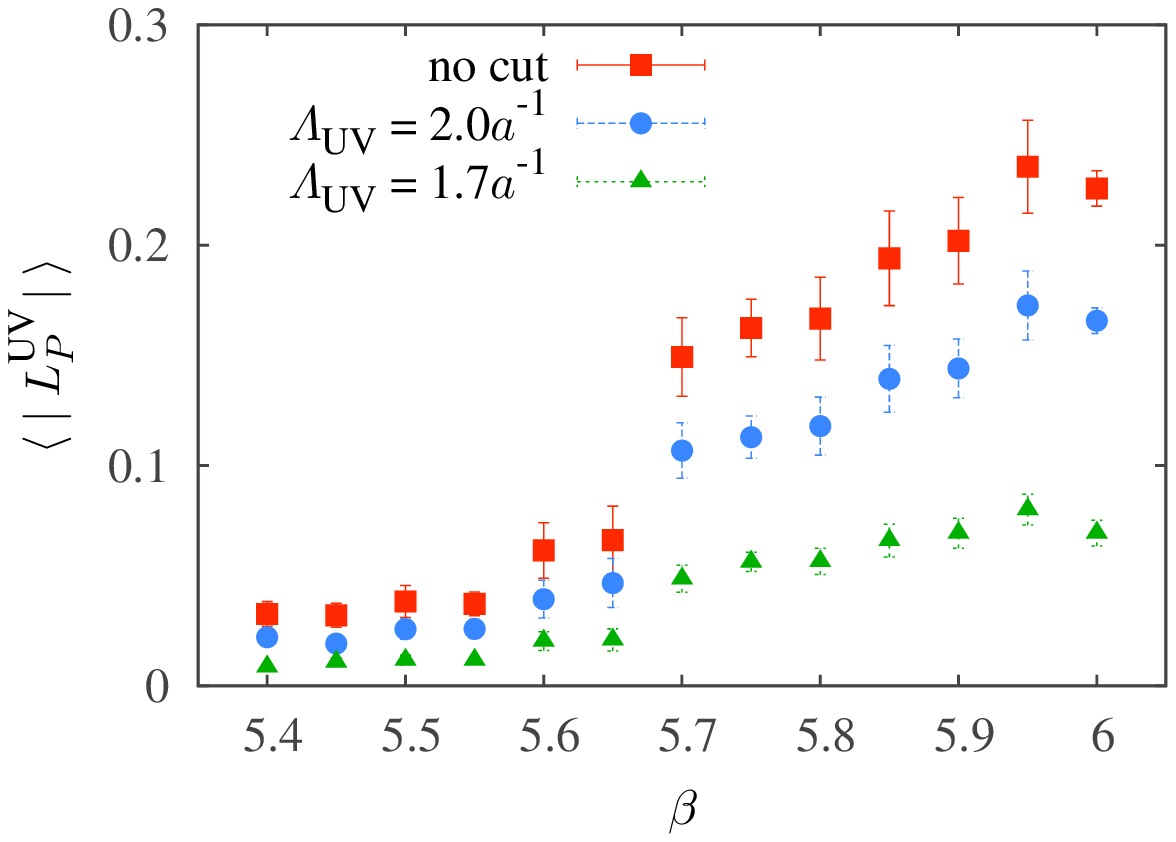}
  \end{center}
  \caption{ \label{fig:PolyakovDiracCut}
  The $\beta$-dependence of the IR/UV-cut Polyakov loop 
  $\langle| L_P^{\rm IR/UV}|\rangle$ 
  on $6^3 \times 4$ lattice.
  (a) $\langle |L_P^{\rm IR}|\rangle $
  for the IR Dirac-mode cut of 
  $\Lambda_{\rm IR} = 0.5a^{-1}$ and $1.0a^{-1}$.
  (b) $\langle| L_P^{\rm UV}|\rangle$ 
  for the UV-cut of 
  $\Lambda_{\rm UV} = 2.0a^{-1}$ and $1.7a^{-1}$.
  According to the mode cut, there appears 
  a reduction factor for $\langle L_P\rangle_{\rm IR/UV}$. 
}
\end{figure}

Figure \ref{fig:PolyakovDiracCut} shows 
$\beta$-dependence of the absolute value of 
the IR-cut Polyakov loop
$\langle |L_P^{\rm IR}| \rangle$ 
with the low-lying cut ($\Lambda_{\rm IR} = 0.5a^{-1},1.0a^{-1}$), and 
the UV-cut Polyakov loop $\langle | L_P^{\rm UV}|\rangle$
with the UV cut ($\Lambda_{\rm UV} = 2.0a^{-1}, 1.7a^{-1}$). 
The numbers of the removed Dirac modes 
for $\Lambda_{\rm IR} = 0.5a^{-1}$ and $1.0a^{-1}$ 
are approximately equal to 
$\Lambda_{\rm UV} = 2.0a^{-1}$ and $1.7a^{-1}$, respectively. 
For comparison, 
we also add the original Polyakov loop $ \langle | L_P | \rangle$, 
which shows the deconfinement phase transition 
around $\beta = 5.6 \sim 5.7$.

As shown in Fig.~\ref{fig:PolyakovDiracCut}, 
both IR-cut and UV-cut Polyakov loops $\langle| L_P^{\rm IR/UV}|\rangle$ 
show almost the same $\beta$-dependence
of the original one $\langle |L_P|\rangle$, 
apart from a normalization factor. 
Thus, we find again no direct connection between 
the Polyakov-loop properties and the Dirac-eigenmodes.
This result is consistent with the similar analysis for the Wilson loop 
using the Dirac-mode expansion method. 
Even after removing IR/UV Dirac-modes, 
the Wilson loop $\langle W \rangle_{\rm IR/UV}$ 
exhibits the area law with the same slope, i.e., the confining force $\sigma$
\cite{Suganuma:2011,Gongyo:2012}.

We also show the $\beta$-dependence 
of the chiral condensate $\langle \bar{q}q \rangle$ 
in the case of removing IR and UV Dirac modes, respectively.
Note that, once the Dirac eigenvalues $\lambda_n$ are obtained, 
the chiral condensate can be easily calculated. 
In fact, the chiral condensate is expressed as 
\begin{eqnarray}
  \langle \bar{q}q \rangle &=& 
  - \frac{1}{V}\mathrm{Tr} \frac{1}{\Slash{D} + m}
  = - \frac{1}{V}\sum_n \frac{1}{i\lambda_n + m} \nonumber \\
  &=& - \frac{1}{V}\left(\sum_{\lambda_n > 0} \frac{2m}{\lambda_n^2 + m^2} - \frac{\nu}{m^2}\right),
\end{eqnarray}
with the total number $\nu$ of the Dirac zero-modes.
Then, the IR/UV-cut chiral condensate is expressed as
\begin{eqnarray}
  \langle \bar{q} q \rangle_{{\rm IR}} &=& - 
  \frac{1}{V}\sum_{\lambda \geq \Lambda_{\rm IR}} \frac{2m}{\lambda^2 + m^2}, \nonumber \\
  \langle \bar{q} q \rangle_{{\rm UV}} &=& - 
  \frac{1}{V}\sum_{0 < \lambda \leq \Lambda_{\rm UV}} \frac{2m}{\lambda^2 + m^2},
  \label{eq-chiral-condensate-cut}
\end{eqnarray}
with the Dirac-mode cut $\Lambda_{\rm IR/UV}$,
apart from the zero-mode contribution.

\begin{figure}
  \begin{center}
    \includegraphics[width=0.85\textwidth,clip]{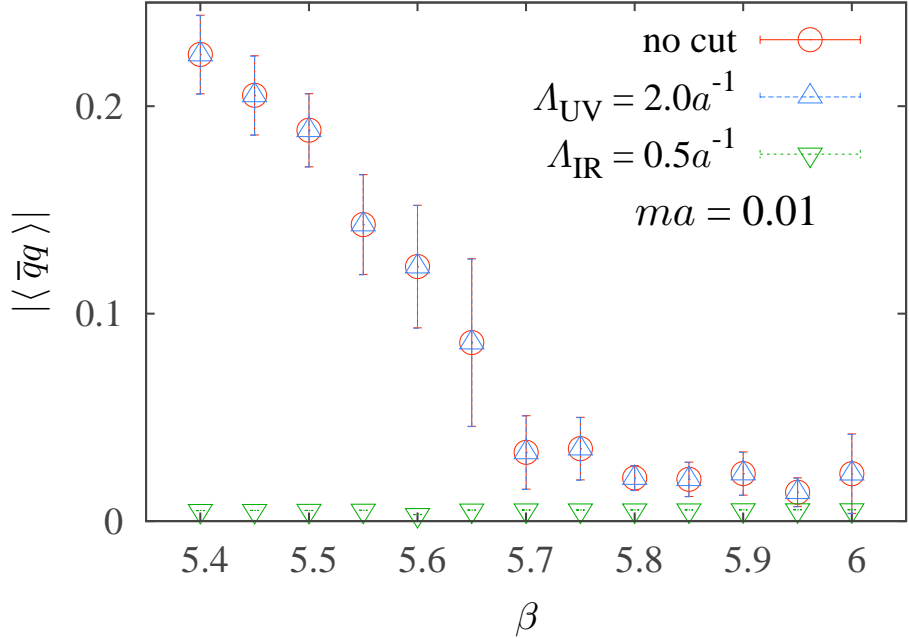}
  \end{center}
  \caption{ \label{fig:ChiralDiracCut}
  The $\beta$-dependence of the chiral condensate 
  $\langle \bar{q}q \rangle_{{\rm IR}/{\rm UV}}$ 
  after removing IR/UV Dirac-modes on $6^4 \times 4$ lattice.
  For comparison, we add the original (no Dirac-mode cut) condensate 
  $\langle \bar{q}q \rangle$, 
  which almost coincides with $\langle \bar{q}q \rangle_{{\rm UV}}$.
}
\end{figure}

Figure \ref{fig:ChiralDiracCut} shows 
the IR-cut chiral condensate 
$\langle \bar{q}q \rangle_{\rm IR}$ 
with $\Lambda_{\rm IR} = 0.5a^{-1}$, 
and the UV-cut chiral condensate 
$\langle \bar{q}q \rangle_{\rm UV}$ 
with $\Lambda_{\rm UV} = 2.0a^{-1}$, 
as a function of $\beta$.
Here, the current quark mass is taken as $m = 0.01a^{-1}$.
For comparison, we also add the original 
(no Dirac-mode cut) chiral condensate
$\langle \bar{q}q \rangle$.
The chiral phase transition occurs around $\beta = 5.6 \sim 5.7$, 
which coincides with the deconfinement transition indicated by 
the Polyakov loop in Fig.~\ref{fig:PolyakovDiracCut}.
The chiral condensate is almost unchanged by the UV-mode cut 
as $\langle \bar{q}q \rangle_{\rm UV} \simeq \langle \bar{q}q \rangle$.
On the other hand, 
the chiral condensate is drastically changed and becomes almost zero as 
$\langle \bar{q}q \rangle_{\rm IR}\simeq 0$ 
by the IR Dirac-mode cut in the whole region of $\beta$.
This clearly shows the essential role of 
the low-lying Dirac-modes to the chiral condensate.
However, the Polyakov-loop behavior is insensitive to the Dirac-mode, 
as shown in Fig.~\ref{fig:PolyakovDiracCut}.

\section{A new method to remove low-lying Dirac-modes from Polyakov loop 
for large lattices
}

In this section, as a convenient formalism, 
we propose a new method to remove low-lying Dirac-modes 
from the Polyakov loop without evaluating full Dirac-modes.
Here, we consider the removal of 
a small number of low-lying Dirac modes, 
since only these modes are responsible to chiral symmetry breaking. 
For the Polyakov loop, unlike the Wilson loop, 
we can easily perform its practical calculation after removing 
the low-lying Dirac modes, by the reformulation 
with respect to the removed IR Dirac-mode space, 
which enables us to calculate with larger lattices. 

As a numerical problem, 
it costs huge computational power to obtain 
the full eigenmodes of the large matrix $\Slash D$, 
and thus our analysis was restricted to 
relatively small lattices in the previous section.
However, in usual eigenvalue problems, 
e.g., in the quantum mechanics, 
one often needs only a small number of low-lying eigenmodes. 
and there are several useful algorithms 
such as the Lanczos method 
to evaluate only low-lying eigenmodes, 
without performing full diagonalization of the matrix.

\subsection{Reformulation of IR Dirac mode subtraction}

The basic idea is to use only the low-lying Dirac modes.
In fact, we calculate only the low-lying Dirac eigenfunction 
$\psi_n(x) \equiv \langle x|n \rangle$ 
for $|\lambda_{n}| < \Lambda_{\rm IR}$, 
and the IR matrix elements 
\begin{eqnarray}
  \langle n|\hat U_\mu|m\rangle
  =\sum_x \psi_n^\dagger(x) U_\mu(x) \psi_m(x+\hat \mu)
\end{eqnarray}
for $|\lambda_{n}|, |\lambda_{m}| < \Lambda_{\rm IR}$. 
We reformulate the Dirac-mode projection 
only with the small number of the low-lying Dirac modes 
of $|\lambda_{n}| < \Lambda_{\rm IR}$.

Here, the IR mode-cut operator $\hat{P}_{\rm IR}$ is expressed as
\begin{eqnarray}
  \hat{P}_{\rm IR} &\equiv&
  \sum_{|\lambda_{n}| \geq \Lambda_{\rm IR}} | n \rangle \langle n| 
  =  1 - \sum_{|\lambda_{n}| < \Lambda_{\rm IR}} | n \rangle \langle n |
  =1-\hat Q,
  \label{eq:P-IR-modified}
\end{eqnarray}
with the IR Dirac-mode projection operator 
\begin{eqnarray}
  \hat Q \equiv \sum_{|\lambda_{n}| < \Lambda_{\rm IR}} | n \rangle \langle n |,
\end{eqnarray} 
corresponding to the low-lying Dirac modes to be removed.
Note that $\sum_{|\lambda_{n}| < \Lambda_{\rm IR}}$ in $\hat Q$
is the sum over only the low-lying modes, 
of which number is small, 
so that this sum is practically performed 
even for larger lattices. 
Then, we reformulate the Dirac-mode projection 
with respect to $\hat Q$ or the small-number sum of 
$\sum_{|\lambda_{n}| < \Lambda_{\rm IR}}$.

We rewrite the IR Dirac-mode cut Polyakov loop as 
\begin{eqnarray}
  \langle L_P\rangle_{\rm IR}
  &=&\frac{1}{3V} \mathrm{Tr}\{ (\hat U_4^P)^{N_t}\}
  =\frac{1}{3V} \mathrm{Tr}\{(\hat U_4 \hat P)^{N_t}\} 
  =\frac{1}{3V} \mathrm{Tr}[\{\hat U_4 (1-\hat Q)\}^{N_t}],
\end{eqnarray}
and expand $\langle L_P\rangle_{\rm IR}$ in terms of $\hat Q$.

As a simple example of the $N_t = 2$ case, 
the IR-cut Polyakov loop $\langle L_P\rangle_{\rm IR}$
is written as 
\begin{eqnarray}
  3V \langle L_P\rangle_{\rm IR}
  &=& \mathrm{Tr}\{\hat U_4 (1-\hat Q) \hat U_4 (1-\hat Q) \}
  \nonumber \\
  &=& \mathrm{Tr}(\hat U_4^2)
  -2 \mathrm{Tr}(\hat Q \hat U_4^2)
  +  \mathrm{Tr}(\hat Q \hat U_4 \hat Q \hat U_4) \nonumber \\
  &=& 3V \langle L_P \rangle
  -2 \sum_{|\lambda_{n}| < \Lambda_{\rm IR}}
  \langle n | \hat{U}_4^2 | n \rangle 
  + \sum_{|\lambda_{n}|, |\lambda_{m}| < \Lambda_{\rm IR}}
  \langle n | \hat{U}_4 | m \rangle
  \langle m | \hat{U}_4 | n \rangle,
  \label{eq:Polyakov-IR-only-cut}
\end{eqnarray}
where $\langle L_P \rangle$
is the ordinary (no cut) Polyakov loop, 
and is easily obtained. 
In Eq.~(\ref{eq:Polyakov-IR-only-cut}), 
we only need the IR matrix elements 
$\langle n | \hat{U}_4 | m \rangle$ and 
\begin{eqnarray}
  \langle n | \hat{U}_4^2 | m \rangle
  &\equiv& \sum_x \sum_y \sum_z \langle n | x \rangle 
  \langle x | \hat{U}_4 | y \rangle
  \langle y | \hat{U}_4 | z \rangle
  \langle z | m \rangle \nonumber \\
  &=& \sum_x \psi_n^\dagger(x) U_4(x) U_4(x+\hat{t}) \psi_m(x+2 \hat{t})
  \quad \qquad
\end{eqnarray}
for $|\lambda_{n}|, |\lambda_{m}| <\Lambda_{\rm IR}$.
Here, $\hat t$  denotes the temporal unit vector in the lattice unit.
In this way, using Eq.~(\ref{eq:Polyakov-IR-only-cut}),
we can remove the contribution of the low-lying Dirac-modes 
from the Polyakov loop, 
only with the IR matrix elements.

For the $N_t = 4$ case, the IR-cut Polyakov loop 
$\langle L_{P}\rangle_{\rm IR}$ is expressed as
  \begin{eqnarray}
    3V \langle L_P\rangle_{\rm IR}
    &=& \mathrm{Tr}\{(\hat U_4^P)^4 \} \nonumber \\
    &=&  \mathrm{Tr}\{\hat U_4 (1-\hat Q) \hat U_4 (1-\hat Q) 
    \hat U_4 (1-\hat Q) \hat U_4 (1-\hat Q) \}
    \nonumber \\
    &=& \mathrm{Tr}(\hat U_4^4) 
    -4\mathrm{Tr}(\hat Q \hat U_4^4)
    +4\mathrm{Tr}(\hat Q \hat U_4   \hat Q \hat U_4^3) \nonumber \\
    &&\quad +2\mathrm{Tr}(\hat Q \hat U_4^2 \hat Q \hat U_4^2)
    -4\mathrm{Tr}(\hat Q \hat U_4 \hat Q \hat U_4 \hat Q \hat U_4^2) 
    +\mathrm{Tr}(\hat Q \hat U_4 \hat Q \hat U_4 \hat Q \hat U_4 \hat Q \hat U_4)
    \nonumber \\
    &=& 3V\left\{ \langle L_P \rangle - L_P^{(1)} + L_P^{(2)}
    - L_P^{(3)} + L_P^{(4)} \right\},
    \label{eq:Polyakov-IR-only-cut4}
  \end{eqnarray}
where $L_P^{(i)}$ ($i=1,2,3,4$) are the IR Dirac-mode contributions 
expanded in terms of $\hat Q$, and are given by
\begin{equation}
  L_P^{(1)} \equiv \frac{4}{3V} \mathrm{Tr} (\hat{Q}\hat{U}_4^4) 
= \frac{4}{3V} \sum_{n_1}^{\rm IR} 
    \langle n_1 | \hat{U}_4^4 | n_1 \rangle, 
  \label{eq:Polyakov-IR-1st}
\end{equation}
\begin{eqnarray}
  L_P^{(2)} &\equiv& 
  \frac{4}{3V}\mathrm{Tr}(\hat{Q}\hat{U}_4 \hat{Q} \hat{U}_4^3)
  + \frac{2}{3V}\mathrm{Tr}(\hat{Q}\hat{U}_4^2 \hat{Q} \hat{U}_4^2) \nonumber  \\
  &=& 
  \frac{4}{3V} \sum_{n_1,n_2}^{\rm IR}
    \langle n_1 | \hat{U}_4 | n_2 \rangle 
    \langle n_2 | \hat{U}_4^3 | n_1 \rangle 
    + \frac{2}{3V} \sum_{n_1,n_2}^{\rm IR}
    \langle n_1 | \hat{U}_4^2 | n_2 \rangle 
    \langle n_2 | \hat{U}_4^2 | n_1 \rangle, 
  \label{eq:Polyakov-IR-2nd}
\end{eqnarray}
\begin{eqnarray}
  L_P^{(3)} &\equiv& \frac{4}{3V}\mathrm{Tr} (\hat{Q} \hat{U}_4\hat{Q}\hat{U}_4 \hat{Q}
  \hat{U}_4^2) 
  = \sum_{n_1,n_2,n_3}^{\rm IR}
    \langle n_1 | \hat{U}_4 | n_2 \rangle \langle n_2 | \hat{U}_4 | n_3 \rangle
    \langle n_3 | \hat{U}_4^2 | n_1 \rangle,  
  \label{eq:Polyakov-IR-3rd}
\end{eqnarray}
  \begin{eqnarray}
    L_P^{(4)} &\equiv& \frac{1}{3V}\mathrm{Tr}
    (\hat{Q}\hat{U}_4\hat{Q}\hat{U}_4\hat{Q}\hat{U}_4\hat{Q}\hat{U}_4) \nonumber \\
    &=& 
  \frac{1}{3V}\sum_{n_1,n_2,n_3,n_4}^{\rm IR}
    \langle n_1 | \hat{U}_4 | n_2 \rangle
    \langle n_2 | \hat{U}_4 | n_3 \rangle 
    \langle n_3 | \hat{U}_4 | n_4 \rangle
    \langle n_4 | \hat{U}_4 | n_1 \rangle.
  \label{eq:Polyakov-IR-4th}
\end{eqnarray}
Here, the summation $\sum^{\rm IR}$ is taken over only 
low-lying Dirac modes with $|\lambda_n| < \Lambda_{\rm IR}$, 
of which number is small. 
In Eq.~(\ref{eq:Polyakov-IR-only-cut4}), 
we only need the IR matrix elements 
$\langle n|\hat U_4^k|m\rangle$ ($k$=1,2,3,4) 
for $|\lambda_{n}|, |\lambda_{m}| <\Lambda_{\rm IR}$, 
and they can be calculated as 
  \begin{eqnarray}
    \langle n | \hat{U}_4^k | m \rangle
    &=&\sum_x \langle n |x \rangle \langle x| \hat{U}_4 |x+\hat t \rangle 
    \langle x+\hat t| \hat{U}_4 |x+2 \hat t \rangle \cdots
    \langle x+(k-1)\hat t| \hat{U}_4 |x+k \hat t \rangle 
    \langle x+k\hat t |m \rangle \nonumber \\
    &=&\sum_x \psi_n^\dagger(x) U_4(x) U_4(x+\hat t) \cdots U_4(x+(k-1)\hat t)
    \psi_m(x+k\hat t).
    \label{eq:matrix-element4}
  \end{eqnarray}
In particular of $k=N_t$, this matrix element is simplified as 
\begin{eqnarray}
  \langle n | \hat{U}_4^{N_t} | m \rangle
  =\sum_x \psi_n^\dagger(x) U_4(x) \cdots U_4(x+(N_t-1)\hat t)
  \psi_m(x) 
  =\sum_x \psi_n^\dagger(x) L_P(x)\psi_m(x),
\end{eqnarray}
with the ordinary Polyakov-loop operator $L_P(x)$. 

Thus, using Eqs.~(\ref{eq:Polyakov-IR-only-cut4}) 
and (\ref{eq:matrix-element4}), we can perform the actual calculation of 
the IR Dirac-mode cut Polyakov loop $\langle L_P \rangle_{\rm IR}$, 
only with the IR matrix elements on the low-lying Dirac modes. 
In this method, we need not 
full diagonalization of the Dirac operator, and hence 
the calculation cost is extremely reduced. 

In principle, we can generalize this method 
for larger temporal-size lattice and the Wilson-loop analysis, 
although the number of the terms becomes larger in these cases.

\subsection{Lattice QCD analysis of 
IR Dirac-mode contribution to Polyakov loop}

Before applying this method to larger-volume lattice calculations,
we investigate the IR Dirac-mode contribution 
to the Polyakov loop, 
$L_P^{(i)}$ defined in 
Eqs.~(\ref{eq:Polyakov-IR-1st})$\sim$(\ref{eq:Polyakov-IR-4th}), 
on the periodic lattice of $6^3 \times 4$ at $\beta = 6.0$,
which corresponds to the deconfined phase.

Here, $L_P^{(i)}$ ($i=1,2,3,4$) are the IR Dirac-mode contributions 
expanded in terms of the number of IR projection $\hat Q$, and satisfy 
\begin{equation}
\langle L_P \rangle_{\rm IR}
= \langle L_P \rangle-L_P^{(1)}+L_P^{(2)}-L_P^{(3)}+L_P^{(4)}.
\label{eq:Polyakov-Q-expansion}
\end{equation}
In this expansion, one can identify $\langle L_P \rangle=L_P^{(0)}$, 
since the original Polyakov loop 
$\langle L_P \rangle$ includes no IR projection $\hat Q$.

\begin{figure}
  \centering
  \includegraphics[width=0.48\textwidth,clip]{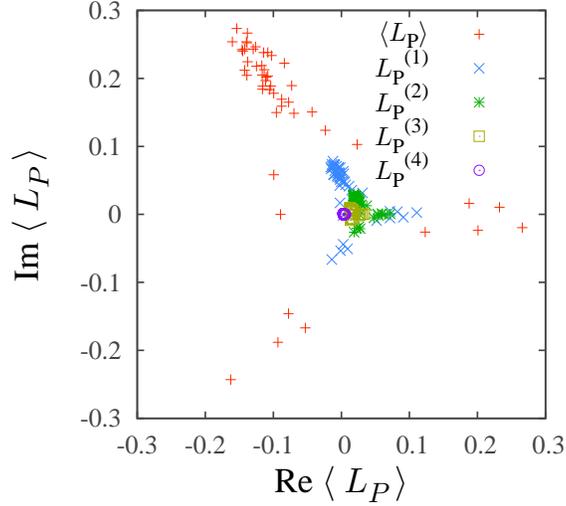}
  \caption{
    \label{fig:IR-contributions}
    IR Dirac-mode contributions to the Polyakov loop, 
    $L_P^{(1)}$, $L_P^{(2)}$, $L_P^{(3)}$, and $L_P^{(4)}$ defined in 
    Eqs.~(\ref{eq:Polyakov-IR-1st})$\sim$(\ref{eq:Polyakov-IR-4th}), 
    in the case of $\Lambda_{\rm IR} = 0.5a^{-1}$,
    in the deconfined phase on the periodic lattice 
    of $6^3 \times 4$ at $\beta = 6.0$.
    For comparison, the original Polyakov loop $\langle L_P \rangle$ is added.
  }
\end{figure}

We show in Fig.~\ref{fig:IR-contributions} 
the scatter plot of 
$L_P^{(1)}$, $L_P^{(2)}$, $L_P^{(3)}$, and $L_P^{(4)}$, 
together with $\langle L_P \rangle$, 
in the case of IR-cut of $\Lambda_{\rm IR} = 0.5a^{-1}$.
As shown in Fig.~\ref{fig:IR-contributions},
all of the IR contributions $L_P^{(i)}$  ($i=1,2,3,4$) are rather small 
in comparison with $\langle L_P \rangle$.
For each gauge configuration, we find 
\begin{equation}
|\langle L_P \rangle| 
\gg |L_P^{(1)}| \gg |L_P^{(2)}| \gg |L_P^{(3)}| \gg |L_P^{(4)}|,
\end{equation}
which leads to $\langle L_P \rangle_{\rm IR} \simeq \langle L_P \rangle$.
Among the IR contribution, 
$L_P^{(1)}$ gives the dominant contribution, 
and higher order terms are almost negligible.
Note also that each $L_P^{(i)}$ distributes in the $Z_3$-center direction 
on the complex plane, and 
$L_P^{(i)}$ in Eq.~(\ref{eq:Polyakov-Q-expansion}) 
partially cancels between odd $i$ and even $i$.
In this way, the approximate magnitude and the $Z_3$ structure of the 
Polyakov loop would be unchanged by the IR Dirac-mode cut.

\subsection{Lattice QCD result for a larger-volume lattice}

\begin{figure}[h]
  \begin{center}
    \includegraphics[width=0.48\textwidth,clip]{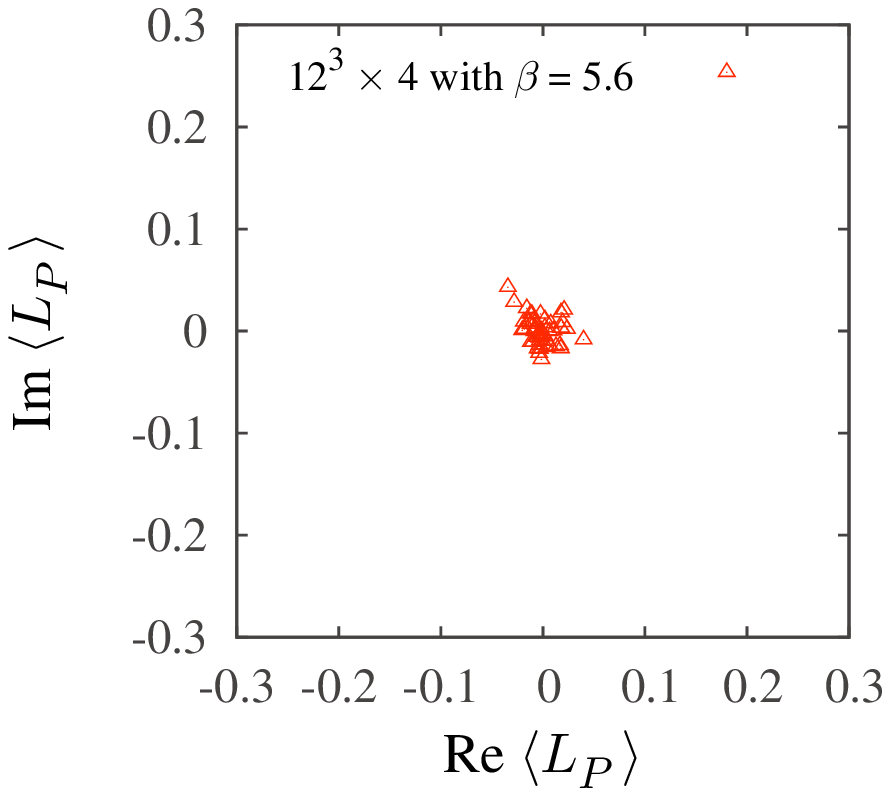}
    \includegraphics[width=0.48\textwidth,clip]{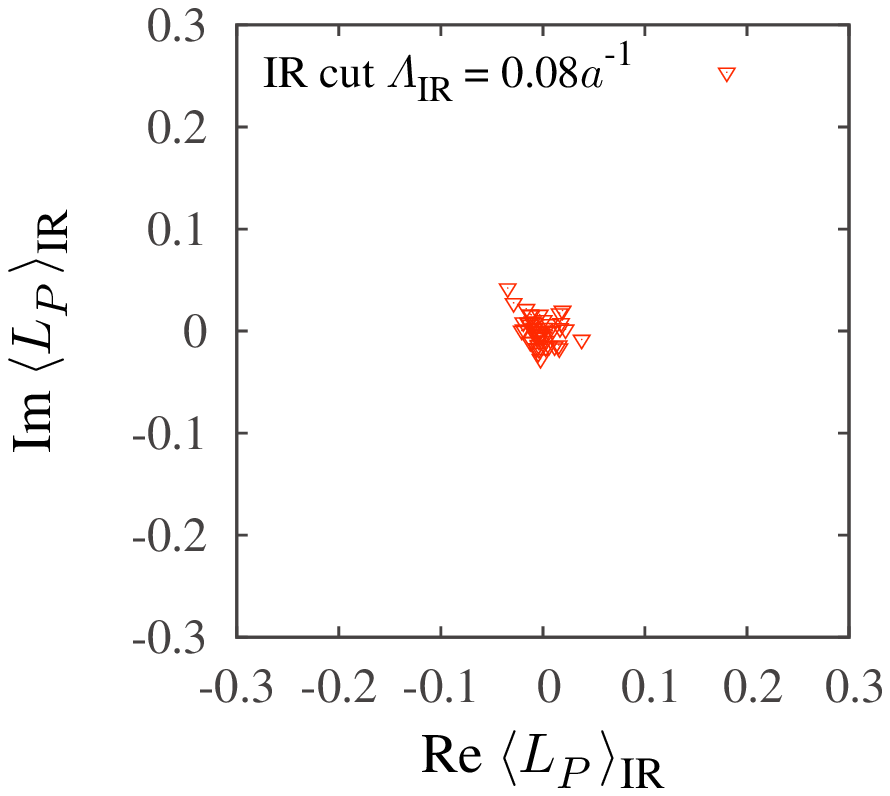}
  \end{center}
  \caption{ \label{fig:LargelatticeConf}
  The Polyakov loop $\langle L_P\rangle$ (upper) and
  the IR Dirac-mode cut Polyakov loop $\langle L_P\rangle_{\rm IR}$ (lower)
  with $\Lambda_{\rm IR} \simeq 0.08a^{-1}$ 
  on $12^3 \times 4$ lattice at $\beta = 5.6$ (confinement phase).
}
\end{figure}

Now, we show the lattice QCD result for 
the Polyakov loop after removing low-lying Dirac modes 
from the confined phase on a larger periodic lattice. 
Figure~\ref{fig:LargelatticeConf} shows 
the scatter plot of the IR-cut Polyakov loop 
$\langle L_P\rangle_{\rm IR}$ 
on the quenched lattice of 
$12^3 \times 4$ at $\beta=5.6$, i.e., $a \simeq$ 0.25~fm and 
$T = 1/(N_ta) \simeq 0.2$~GeV below 
$T_c\simeq 0.26$~GeV, using 50 gauge configurations.
For comparison, the original (no-cut) Polyakov loop 
$\langle L_P \rangle$ is also shown in Fig.~\ref{fig:LargelatticeConf}.
Here, we use ARPACK \cite{ARPACK} to calculate
low-lying Dirac eigenmodes.
On the IR-cut parameter, 
we use $\Lambda_{\rm IR}= 0.08a^{-1}$, which corresponds to 
the removal of about 180 low-lying Dirac modes from the total 20736 modes.
In this case, the IR-cut quark condensate 
$\langle \bar qq\rangle_{\rm IR}$
is reduced to be 
only about 7\%, i.e., 
$\langle \bar qq\rangle_{\rm IR}/\langle \bar qq\rangle \simeq 0.07$, 
around the physical current-quark mass 
of $m \simeq 0.006a^{-1} \simeq$ 5~MeV.

\begin{figure}[h]
  \begin{center}
    \includegraphics[width=0.48\textwidth,clip]{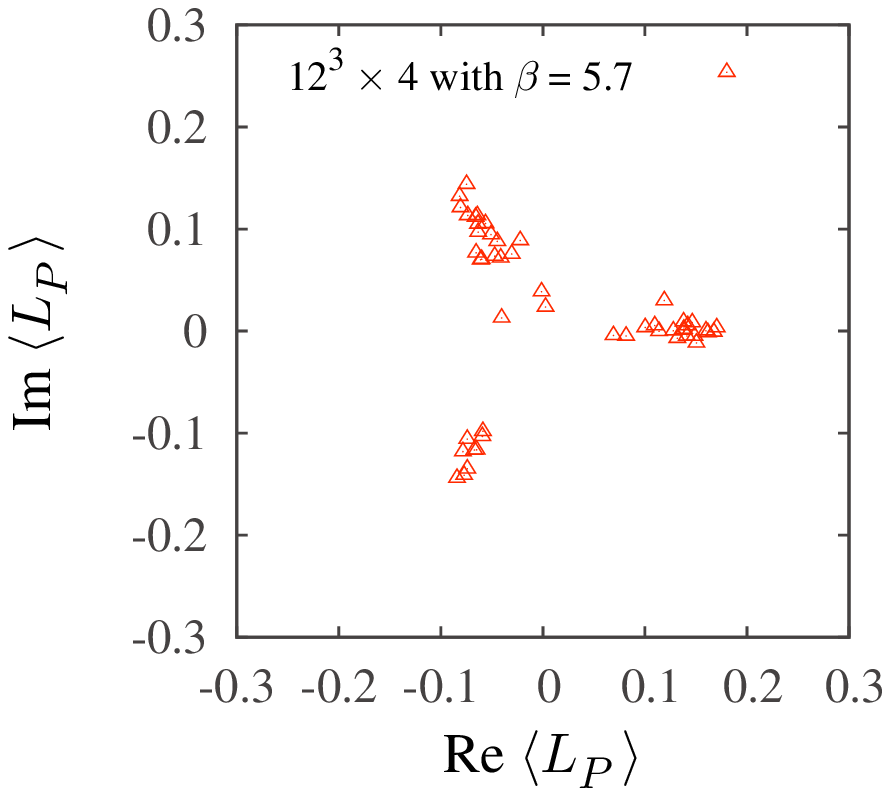}
    \includegraphics[width=0.48\textwidth,clip]{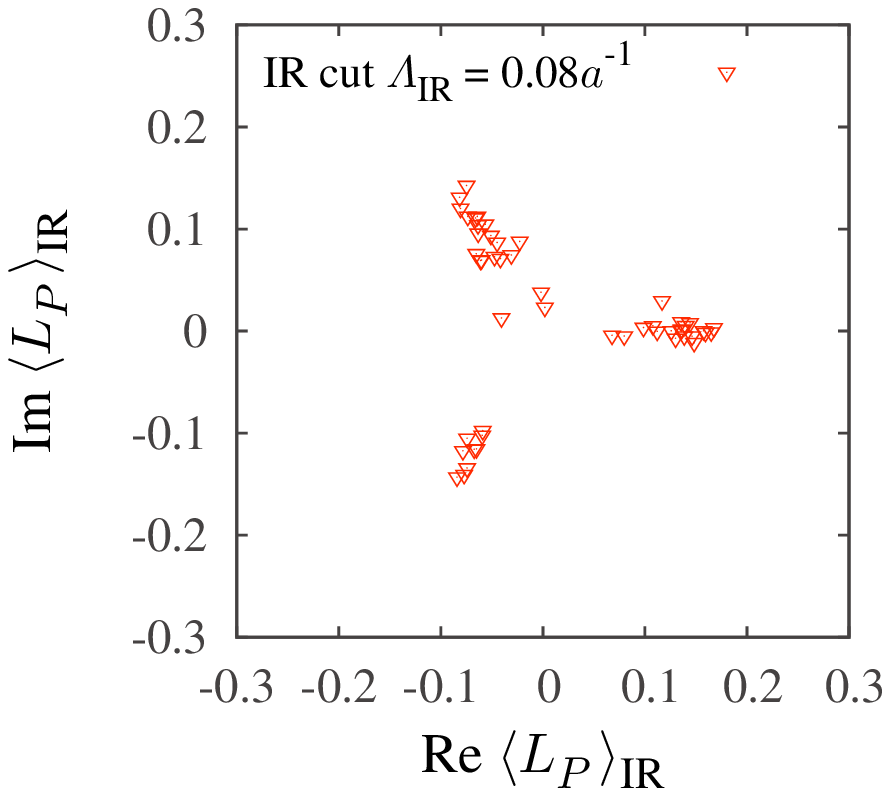}
  \end{center}
  \caption{ \label{fig:LargelatticeDeconf}
  The Polyakov loop $\langle L_P\rangle$ (upper) and
  the IR Dirac-mode cut Polyakov loop $\langle L_P\rangle_{\rm IR}$ (lower)
  with $\Lambda_{\rm IR} \simeq 0.08a^{-1}$ 
  on $12^3 \times 4$ lattice at $\beta = 5.7$ (deconfinement phase).
}
\end{figure}

Note again that the IR-cut Polyakov loop is almost zero as 
$\langle L_P\rangle_{\rm IR} \simeq 0$
and the $Z_3$ center symmetry is kept, that is, 
the confinement is still realized, 
even without the low-lying Dirac modes,
which are essential for chiral symmetry breaking.

Next, we show the removal of low-lying Dirac modes 
from the deconfined phase on a larger periodic lattice. 
Figure~\ref{fig:LargelatticeDeconf} shows 
the IR-cut Polyakov loop $\langle L_P\rangle_{\rm IR}$
together with $\langle L_P \rangle$
on $12^3 \times 4$ at $\beta=5.7$, i.e., 
$a \simeq$ 0.186~fm 
\cite{Takahashi} 
and $T\equiv 1/(N_ta)\simeq$ 0.27~GeV above $T_c$, 
using 50 gauge configurations. 
We use $\Lambda_{\rm IR}= 0.08a^{-1}$, which corresponds to 
the removal of about 120 low-lying Dirac modes from the total 20736 modes.
In this case, we find 
$\langle L_P\rangle_{\rm IR}\simeq \langle L_P \rangle$ 
for each gauge configuration, 
and observe almost no effect of the IR Dirac-mode removal 
for the Polyakov loop.

Thus, for both confined and deconfined phases, 
the Polyakov-loop behavior is almost unchanged 
by removing the low-lying Dirac modes, 
in terms of the zero/non-zero expectation value 
and the $Z_3$ center symmetry.
In fact, we find again the IR Dirac-mode insensitivity 
to the Polyakov-loop or the confinement property 
also for the larger volume lattice.

\section{Summary and Concluding Remarks}

In this paper, we have investigated the direct correspondence 
between the Polyakov loop and the Dirac eigenmodes 
in a gauge-invariant manner 
in SU(3) lattice QCD at the quenched level 
in both confined and deconfined phases.
Based on the Dirac-mode expansion method, 
we have removed the essential ingredient of 
chiral symmetry breaking from the Polyakov loop.

In the confined phase, 
we have found that the IR-cut Polyakov loop 
$\langle L_P\rangle_{\rm IR}$ is still almost zero 
even without low-lying Dirac eigenmodes.
As shown in the Banks-Casher relation,
these low-lying modes are essential for chiral symmetry breaking. 
This result indicates that the system still remains in the confined phase 
after the effective restoration of chiral symmetry.
We have also analyzed the role of high (UV) Dirac-modes,
and have found that the UV-cut Polyakov loop 
$\langle L_P\rangle_{\rm UV}$ is also zero. 
These results indicate that there is no definite Dirac-modes region 
relevant for the Polyakov-loop behavior, in fact, each Dirac eigenmode seems to feel 
that the system is in the confined phase.

This Dirac-mode insensitivity to the confinement 
is consistent with the previous Wilson-loop analysis 
with the Dirac-mode expansion in Refs.\cite{Suganuma:2011,Gongyo:2012}, 
where the Wilson loop shows area law 
and linear confining potential is almost unchanged 
even without low-lying or high Dirac eigenmodes. 
These results are also consistent 
with the existence of hadrons as bound states 
without low-lying Dirac-modes \cite{Lang:2011,Glozman:2012}.
Also, Gattringer's formula suggests that 
the existence of Dirac zero-modes does not seem to contribute 
to the Polyakov loop \cite{Gattringer:2006}.

Next, we have analyzed the Polyakov loop 
in the deconfined phase at high temperature, 
where the Polyakov loop $\langle L_P \rangle$
has a non-zero expectation value, 
and its value distributes in $Z_3$ direction in the complex plane. 
We have found that both IR-cut and UV-cut Polyakov loops 
$\langle L_P\rangle_{\rm IR/UV}$ have 
the same properties of the non-zero expectation value 
and the $Z_3$ symmetry breaking.

We have also investigated the temperature dependence 
of the IR/UV-cut Polyakov loop $\langle L_P\rangle_{\rm IR/UV}$, 
and have found that $\langle L_P\rangle_{\rm IR/UV}$ 
shows almost the same temperature dependence 
as the original Polyakov loop $\langle L_P \rangle$,
while the IR-cut chiral condensate 
$\langle \bar qq \rangle_{\rm IR}$ 
becomes almost zero even below $T_c$, 
after removing the low-lying Dirac-modes.

Finally, we have 
developed a new method to calculate 
the IR-cut Polyakov loop $\langle L_P\rangle_{\rm IR}$ 
in a larger volume at finite temperature, 
by the reformulation with respect to the removed IR Dirac-mode space, 
and have found again the IR Dirac-mode insensitivity to 
the Polyakov loop or the confinement property 
on a larger lattice of $12^3\times 4$. 

These lattice QCD results and related studies 
\cite{Lang:2011,Glozman:2012,Suganuma:2011,Gongyo:2012,TMDoi}
suggest that each eigenmode 
has the information of confinement/deconfinement,
i.e., the ``seed'' of confinement
is distributed in a wider region of the Dirac eigenmodes.
We consider that there is no direct connection 
between color confinement and chiral symmetry breaking
through the Dirac eigenmodes.
In fact, the one-to-one correspondence 
does not hold between confinement and chiral symmetry breaking in QCD, 
and their appearance can be different in QCD.
This mismatch may suggest 
richer QCD phenomena and richer structures in QCD phase diagram.
It is interesting to proceed full QCD and investigate 
dynamical quark effects in our framework.
It is also interesting to search 
the relevant modes only for color confinement 
but irrelevant for chiral symmetry breaking \cite{Iritani:2012FP}.

\section*{Acknowledgements}
  The authors thank Shinya Gongyo for his contribution to 
  the early stage of this study.
  This work is in part supported by the Grant for Scientific Research 
  [(C) No.23540306, Priority Areas ``New Hadrons'' (E01:21105006)] and 
  a Grant-in-Aid for JSPS Fellows [No. 23-752] from the Ministry of Education, 
  Culture, Science and Technology (MEXT) of Japan.
  The lattice QCD calculations have been done on NEC-SX8R and NEC-SX9 at 
  Osaka University.

\appendix
\section{Intermediate Dirac-mode removal for Polyakov loop}
In this appendix, we study the role of the intermediate (IM) Dirac-modes 
to the Polyakov loop in both confined and deconfined phases. 
We consider the cut of IM Dirac modes of $\Lambda_1 < |\lambda_n| < \Lambda_2$.
Then, the IM-cut Polyakov-loop is defined as
\begin{equation}
  \langle L_P\rangle_{\rm IM} \equiv \frac{1}{3V}
  \mathrm{tr} 
  \sum_{|\lambda_{n_i}| \le \Lambda_{\rm 1}, \\ 
  \Lambda_{\rm 2} \le |\lambda_{n_i}|}
  \langle n_1 | \hat{U}_4 | n_2 \rangle \cdots
  \langle n_{N_t} | \hat{U}_4 | n_1 \rangle,
  \label{eq:PolyakovIMcut}
\end{equation}
with the cut parameters, $\Lambda_1$ and $\Lambda_2$.

\begin{figure*}[h]
  \centering
  \includegraphics[width=0.32\textwidth,clip]{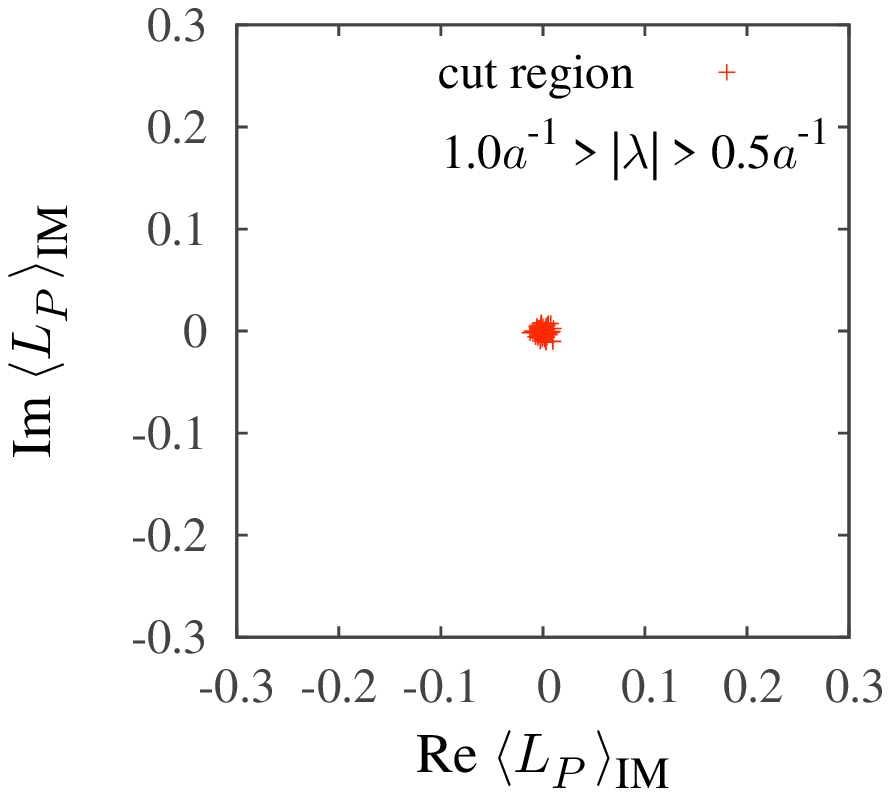}
  \includegraphics[width=0.32\textwidth,clip]{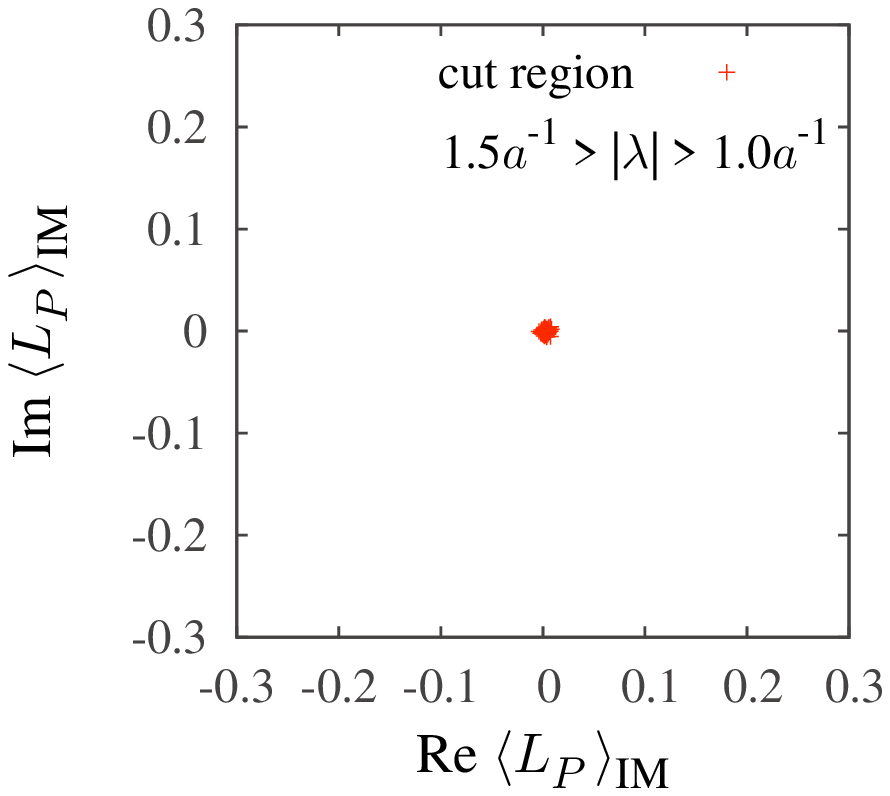}
  \includegraphics[width=0.32\textwidth,clip]{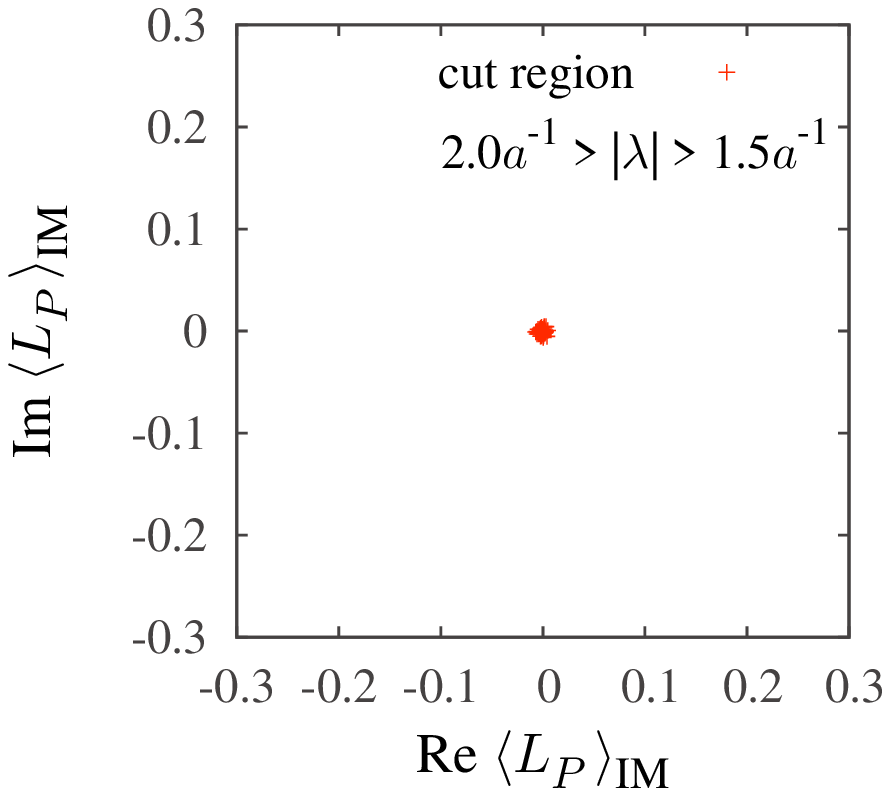}
  \caption{\label{fig:PolyakovScatterConfinedIM}
  The IM-cut Polyakov loop on the periodic lattice of $6^4$ 
  at $\beta = 5.6$ in the confined phase. 
  The cut region of the Dirac mode is 
  (a) $|\lambda| \in (0.5a^{-1},1.0a^{-1})$,
  (b) $|\lambda| \in (1.0a^{-1},1.5a^{-1})$, and 
  (c) $|\lambda| \in (1.5a^{-1},2.0a^{-1})$, respectively.
}
  \end{figure*}

\begin{figure*}[h]
    \centering
    \includegraphics[width=0.32\textwidth,clip]{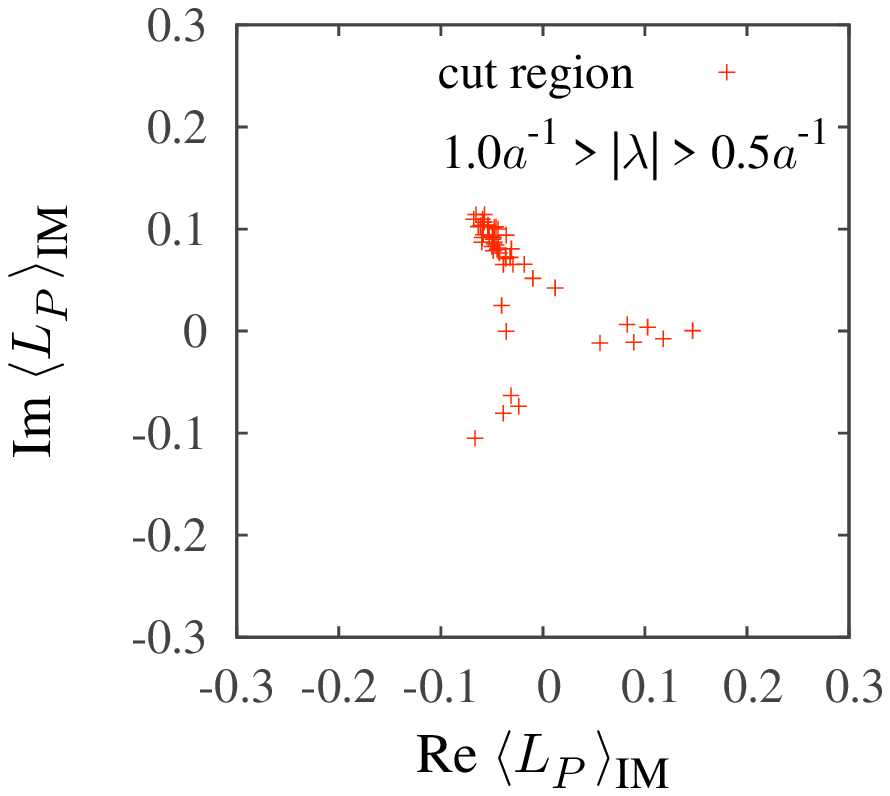}
    \includegraphics[width=0.32\textwidth,clip]{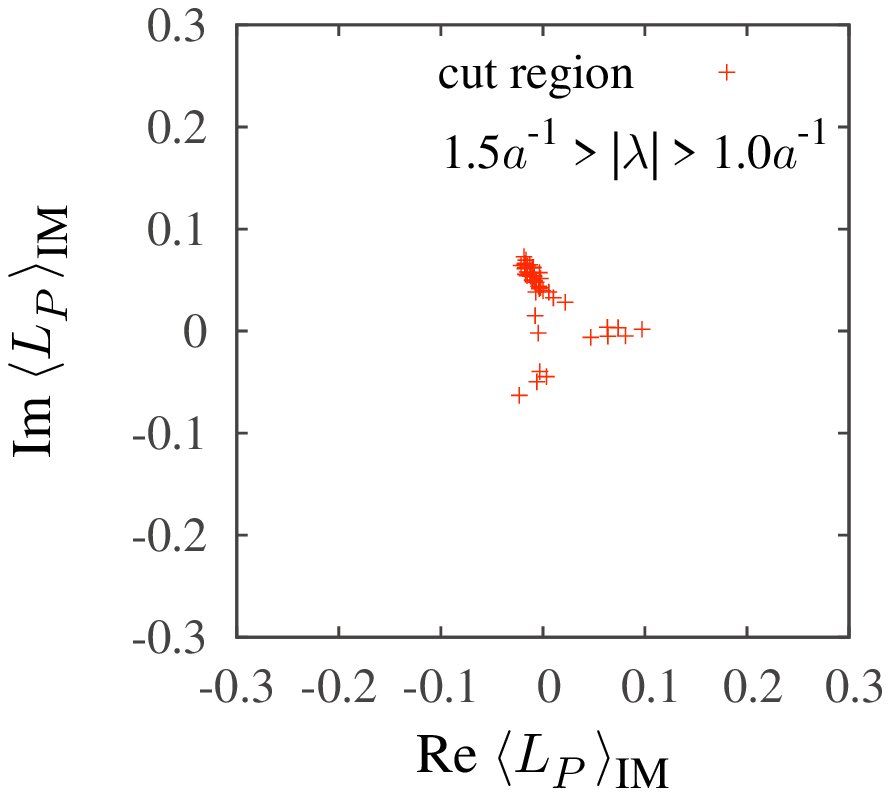}
    \includegraphics[width=0.32\textwidth,clip]{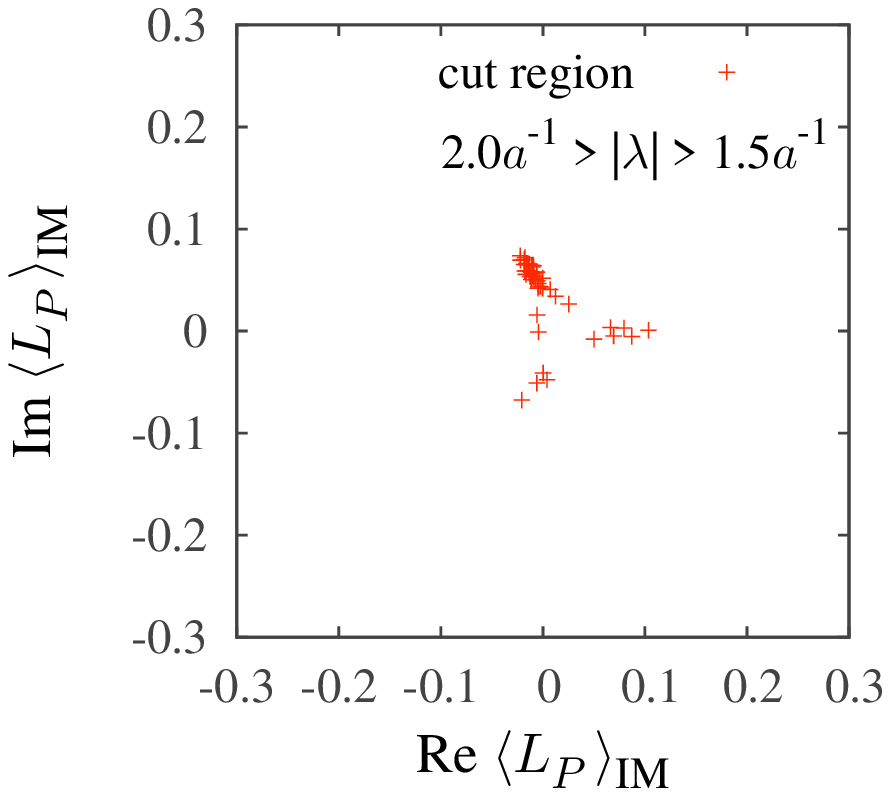}
    \caption{\label{fig:PolyakovScatterDeconfinedIM}
    The IM-cut Polyakov loop on the periodic lattice of $6^3 \times 4$ 
    at $\beta = 6.0$ in the deconfined phase. 
    The cut region of the Dirac mode is 
    (a) $|\lambda| \in (0.5a^{-1},1.0a^{-1})$,
    (b) $|\lambda| \in (1.0a^{-1},1.5a^{-1})$, and 
    (c) $|\lambda| \in (1.5a^{-1},2.0a^{-1})$, respectively.
  }
\end{figure*}

Figures \ref{fig:PolyakovScatterConfinedIM} and 
\ref{fig:PolyakovScatterDeconfinedIM} 
show the IM-cut Polyakov loop $\langle L_P\rangle_{\rm IM}$
on the periodic lattice of $6^4$ at $\beta = 5.6$ in the confined phase, 
and that of $6^3 \times 4$ at $\beta = 6.0$ in the deconfined phase, 
respectively. 
Here, we remove the IM modes of $0.5-1.0[a^{-1}]$, $1.0-1.5[a^{-1}]$,
and $1.5-2.0[a^{-1}]$, respectively.


In the confined phase, the IM-cut Polyakov loop 
$\langle L_P\rangle_{\rm IM}$ is almost zero, 
and $\langle L_P\rangle_{\rm IM}$ has non-zero expectation value 
in the deconfined phase.
These Dirac-mode insensitivities are similar to 
the case of IR/UV-cut Polyakov loops.

\end{document}